\documentclass{aa}
\usepackage{graphicx}
\usepackage{txfonts}
\usepackage{hyperref}
\begin{document} 

   \title{Characterization of the K2-38 planetary system.   
   \thanks{Based (in part) on Guaranteed Time Observations collected at the European Southern Observatory under ESO programmes1102.C-0744, 112.C-0958, and 1104.C-0350 by the ESPRESSO Consortium.}
   \thanks{The ESPRESSO RVs used in this paper are available in electronic form at the CDS via anonymous ftp to \url{cdsarc.u-strasbg.fr} (130.79.128.5) or via \url{http://cdsweb.u-strasbg.fr/cgi-bin/qcat?J/A+A/}}
   }

   \subtitle{Unraveling one of the densest planets known to date.}

   \author{B. Toledo-Padr\'on\inst{1,2} \thanks{E-mail: btoledo@iac.es}, 
          C. Lovis\inst{3}, 
          A. Su\'arez Mascare\~no\inst{1,2}, 
          S. C. C. Barros\inst{4,5}, 
          J. I. Gonz\'alez Hern\'andez\inst{1,2}, 
          A. Sozzetti\inst{6}, 
          F. Bouchy\inst{3}, 
          M. R. Zapatero Osorio\inst{7}, 
          R. Rebolo\inst{1,2,8}, 
          S. Cristiani\inst{9}, 
          F. A. Pepe\inst{3}, 
		  N. C. Santos\inst{4,5}, 
		  S. G. Sousa\inst{4}, 
		  H. M. Tabernero\inst{4}, 
		  J. Lillo-Box\inst{10}, 
		  D. Bossini\inst{4}, 
		  V. Adibekyan\inst{4,5}, 
		  R. Allart\inst{3}, 
		  M. Damasso\inst{6}, 
		  V. D'Odorico\inst{9,11}, 
		  P. Figueira\inst{12,4}, 
		  B. Lavie\inst{3}, 
		  G. Lo Curto\inst{13}, 
		  A. Mehner\inst{13}, 
		  G. Micela\inst{14}, 
		  A. Modigliani\inst{13}, 
		  N. J. Nunes\inst{15,16}, 
		  E. Pall\'e\inst{1,2}, 
          M. Abreu\inst{15,16}, 
		  M. Affolter\inst{17}, 
		  Y. Alibert\inst{17}, 
		  M. Aliverti\inst{18}, 
		  C. Allende Prieto\inst{1,2}, 
		  D. Alves\inst{15,16}, 
		  M. Amate\inst{1}, 
		  G. Avila\inst{13}, 
		  V. Baldini\inst{9}, 
		  T. Bandy\inst{17}, 
		  S. Benatti\inst{19,14}, 
		  W. Benz\inst{17}, 
		  A. Bianco\inst{18}, 
          C. Broeg\inst{17}, 
		  A. Cabral\inst{15,16}, 
		  G. Calderone\inst{9}, 
		  R. Cirami\inst{9}, 
		  J. Coelho\inst{15,16}, 
		  P. Conconi\inst{18}, 
		  I. Coretti\inst{9}, 
		  C. Cumani\inst{13}, 
		  G. Cupani\inst{9}, 
		  S. Deiries\inst{13}, 
		  H. Dekker\inst{13}, 
		  B. Delabre\inst{13}, 
		  O. Demangeon\inst{4,5}, 
		  P. Di Marcantonio\inst{9}, 
		  D. Ehrenreich\inst{3}, 
		  A. Fragoso\inst{1}, 
		  L. Genolet\inst{3}, 
		  M. Genoni\inst{18}, 
		  R. G\'enova Santos\inst{1,2}, 
		  I. Hughes\inst{3}, 
		  O. Iwert\inst{13}, 
		  J. Knudstrup\inst{13}, 
		  M. Landoni\inst{18}, 
		  J. L. Lizon\inst{13}, 
		  C. Maire\inst{3}, 
		  A. Manescau\inst{13}, 
		  C. J. A. P. Martins\inst{4,20}, 
		  D. Mégevand\inst{3}, 
		  P. Molaro\inst{9,21}, 
		  M. J. P. F. G. Monteiro\inst{4,5}, 
          M. A. Monteiro\inst{4}, 
		  M. Moschetti\inst{18}, 
		  E. Mueller\inst{13}, 
		  L. Oggioni\inst{18}, 
		  A. Oliveira\inst{15,16}, 
		  M. Oshagh\inst{1,2}, 
		  G. Pariani\inst{18}, 
		  L. Pasquini\inst{12}, 
		  E. Poretti\inst{18,22}, 
		  J. L. Rasilla\inst{1}, 
		  E. Redaelli\inst{18}, 
		  M. Riva\inst{18}, 
		  S. Santana Tschudi\inst{1,12}, 
		  P. Santin\inst{9}, 
		  P. Santos\inst{15,16}, 
		  A. Segovia\inst{3}, 
		  D. Sosnowska\inst{3}, 
		  P. Spanò\inst{18}, 
		  F. Tenegi\inst{1}, 
		  S. Udry\inst{3}, 
		  A. Zanutta\inst{18}, 
		  and
		  F. Zerbi\inst{18} 
		  \newline
		  \newline
		  \textit{(Affiliations can be found after the references)}
		  }
          
   \titlerunning{Characterization of the K2-38 planetary system}
          
   \authorrunning{B. Toledo-Padr\'on et al.}

   \institute{}

   \date{\centering Accepted A\&A June 2020} 

 
  \abstract
   {An accurate characterization of the known exoplanet population is key to understanding the origin and evolution of planetary systems. Determining true planetary masses through the radial velocity (RV) method is expected to experience a great improvement thanks to the availability of ultra-stable echelle spectrographs.}
   {We took advantage of the extreme precision of the new-generation echelle spectrograph ESPRESSO to characterize the transiting planetary system orbiting the G2V star K2-38 located at 194~pc from the Sun with $V\sim$11.4. This system is particularly interesting because it could contain the densest planet detected to date.}
   {We carried out a photometric analysis of the available K2 photometric light curve of this star to measure the radius of its two known planets, K2-38b and K2-38c, with $P_{\rm  b}$=4.01593$\pm$0.00050~d and $P_{\rm  c}$=10.56103$\pm$0.00090~d, respectively. Using 43 ESPRESSO high-precision radial velocity measurements taken over the course of 8 months along with the 14 previously published HIRES RV measurements, we modeled the orbits of the two planets through a Markov Chain Monte Carlo (MCMC) analysis, significantly improving their mass measurements.}
   {Using ESPRESSO spectra, we derived the stellar parameters, $T_{\rm eff}$=5731$\pm$66, $\log g$=4.38$\pm$0.11~dex, and $[Fe/H]$=0.26$\pm$0.05~dex, and thus the mass and radius of K2-38,  $M_{\star}$=1.03 $^{+0.04}_{-0.02}$~M$_{\oplus}$ and $R_{\star}$=1.06 $^{+0.09}_{-0.06}$~R$_{\oplus}$. We determine new values for the planetary properties of both planets. We characterize K2-38b as a super-Earth with $R_{\rm P}$=1.54$\pm$0.14~R$_{\rm \oplus}$ and $M_{\rm p}$=7.3$^{+1.1}_{-1.0}$~M$_{\oplus}$, and K2-38c as a sub-Neptune with $R_{\rm P}$=2.29$\pm$0.26~R$_{\rm \oplus}$ and $M_{\rm p}$=8.3$^{+1.3}_{-1.3}$~M$_{\oplus}$. Combining the radius and mass measurements, we derived a mean density of $\rho_{\rm p}$=11.0$^{+4.1}_{-2.8}$~g cm$^{-3}$ for K2-38b and $\rho_{\rm p}$=3.8$^{+1.8}_{-1.1}$~g~cm$^{-3}$ for K2-38c, confirming K2-38b as one of the densest planets known to date.}
   {The best description for the composition of K2-38b comes from an iron-rich Mercury-like model, while K2-38c is better described by an rocky model with a H2 envelope. The maximum collision stripping boundary shows how giant impacts could be the cause for the high density of K2-38b. The irradiation received by each planet places them on opposite sides of the radius valley. We find evidence of a long-period signal in the radial velocity time-series whose origin could be linked to a 0.25-3~M$_{\rm J}$ planet or stellar activity.}


   \maketitle
   
\newpage

\section{Introduction}

The total number of known exoplanets keeps increasing every day. With main contributions from the transit and radial velocity methods, the number of detections has already surpassed the 4000 landmark. Despite this promising evolution in terms of detection, the characterization of these objects is a matter that has not been addressed with the same level of success. From the large sample of confirmed extrasolar planets, only $\sim$20\% have the true dynamical mass measured (this fraction increases up to $\sim$40\% when including projected mass $M \sin i$ determinations), and only $\sim$10\% of the complete sample have a published measurement of their density coming from mass and radius values with uncertainties lower than 25\% \footnote{source: \url{https:/exoplanets.nasa.gov/}}. This limits studies about planetary formation or the atmospheric characterization of transiting planets, for which a precise mass measurement is required.

Using the sample of known transiting planets, several studies have been conducted to investigate the mass-radius relation for exoplanets, taking into account the role of the flux received from the star \citep{2012A&A...540A..99E,2012PASP..124..323K,2013ApJ...768...14W,2017ApJ...834...17C,2017A&A...604A..83B,2019A&A...630A.135U} and also their bulk composition \citep{2012ApJ...744...59S,2015A&A...577A..83D,2017A&A...597A..37D}. These studies about the formation and evolution of exoplanetary systems have shown that planet radii increase with the mass in the case of low-mass planets, while these two quantities present a constant or even negative relation for high-mass planets (higher than 100 M$_{\oplus}$). In the lower-mass region, we found the two most common types of planets in our galaxy: super-Earths and sub-Neptunes. One of the most relevant differences that separates these two types of planets in close-in orbits is the existence of radius valley between them \citep{2013ApJ...775..105O,2013ApJ...776....2L,2016ApJ...831..180C,2017AJ....154..109F}. The theoretical models predict evaporation effects on planets under high levels of irradiance from their host star \citep{2003ApJ...598L.121L,2007Icar..187..358H,2011A&A...529A.136E,2012MNRAS.425.2931O}, which has been already constrained in terms of the planetary binding energy \citep{2012ApJ...761...59L,2013ApJ...770..131L,2013ApJ...775..105O,2014ApJ...792....1L}. These effects can turn into atmosphere and mass losses, especially for planets with a short orbital period \citep{2004A&A...419L..13B,2009ApJ...693...23M,2012ApJ...761...59L}. This phenomenon explains the lower frequency of low-density planets at a short distance to their parent star \citep{2013ApJ...775..105O}. Furthermore, a higher incident flux (along with other heating effects) can produce an increase in the planet radii \citep{1996ApJ...459L..35G}. The radius gap indicates that for a certain level of insolation flux, the more massive planets are able to maintain their atmospheres, while the less massive ones lose their gas envelope. This gap could also be caused by other mechanisms such as core-powered atmospheric mass loss \citep{2019MNRAS.487...24G} or in-situ rocky planet formation in a gas-poor environment \citep{2016ApJ...817...90L}.

In this work, we aim to shed more light on these points by studying the solar-type star K2-38 (2MASS J16000805-2311213) using the Echelle SPectrograph for Rocky Exoplanets and Stable Spectroscopic Observations (ESPRESSO). K2-38 is a \textit{V}=11.34 high-proper motion G2-type star \citep{2016yCat.2336....0H} located at 194 pc from the Sun \citep{2018A&A...616A...1G} in the Scorpius constellation. This solar-type star has two detected planets \citep{2016ApJS..226....7C}, preliminarily characterized using 14 HIRES spectra \citep{2016ApJ...827...78S}: the first planet is a high-density super-Earth with an orbital period of 4d, and the second planet is a low-density sub-Neptune with a period of 10.6d. \citet{2016ApJ...827...78S} reported radii of $R_{\rm b}$=1.55$\pm$0.16~R$_{\oplus}$ and $R_{\rm c}$=2.42$\pm$0.29~R$_{\oplus}$, and masses of $M_{\rm b}$=12.0$\pm$2.9~M$_{\oplus}$ and $M_{\rm c}$=9.9$\pm$4.6~M$_{\oplus}$ for K2-38b and K2-38c, respectively. The authors claimed that K2-38b could be the densest planet known to date due to its bulk density of $\rho_{\rm p}=$17.5$^{+8.5}_{-6.2}$~g~cm$^{-3}$. We take advantage of the high RV precision provided by ESPRESSO to improve the mass measurement of the two planets along with other properties. We study how this system fits into the radius valley scenario and explore the mass-radius relation in terms of the incident stellar flux.

The paper is structured into five sections. In Sect.~\ref{sec:Data} we describe our dataset. Sect.~\ref{sec:K2-38} shows the stellar parameter analysis and the chemical abundances study. Sect.~\ref{sec:Analysis} details the photometric and spectroscopic analysis carried out for the planetary characterization. Sect.~\ref{sec:Discussion} contains the discussion of the previous analysis and Sect.~\ref{sec:Conclusions} presents the conclusions.

\section{Data}

\label{sec:Data}

ESPRESSO is an ultra-stable fiber-fed high-resolution spectrograph installed at the Very Large Telescope (VLT) in the Paranal Observatory, Chile \citep{2014AN....335....8P}. We acquired 43 spectra of K2-38 as part of the ESPRESSO Guaranteed Time Observation (programmes 1102.C-0744, 1102.C-0958, and 1104.C-0350) using the instrument in single Unit Telescope mode. This mode provides a spectral resolution of $R$=140\,000 using an on-sky fiber of 1". The time-span of the sample covers about 240d: from 19 February 2019 (BJD=2458533.8) to 16 October 2019 (BJD=2458773.5). We split the dataset into two subsets due to the maintenance operations carried out in the instrument on the last weeks of June 2019 to update its fiber link that produce an offset in the RV measurements (Pepe et al., in preparation). All of these spectra were taken with an exposure time of 900s, except for five cases where we increased it due to bad atmospheric conditions, never exceeding 1800s to allow a precise solar system barycentric RV correction. We obtained a mean signal-to-noise ratio (S/N) per extracted pixel of 56.1 at 550~nm for the whole sample.

We used the public version of the ESPRESSO pipeline Data-Reduction-Software (DRS) \footnote{\url{https://www.eso.org/sci/software/pipelines/espresso/espresso-pipe-recipes.html}} to compute the RVs and several stellar activity indicators. The pipeline provides a cross-correlation function (CCF) for each spectrum using a G2 mask that covers the entire wavelength range of the instrument (between 3800 and 7880~\AA). The CCFs were built using a RV step of 0.5~km~s$^{-1}$ within a range between -55 and -15~km~s$^{-1}$ centered on the systemic velocity of the star. In the RV time-series, we achieved a RV precision of 1.0~m~s$^{-1}$ with a RMS of 3.6~m~s$^{-1}$, an extremely good result for a relatively faint G2 star (\textit{V}=11.34) like K2-38. We combined this time-series with the 14 HIRES RV measurements of K2-38 published by \citet{2016ApJ...827...78S}. This dataset presents a time-span of $\sim$100d: from 24 June 2015 (BJD=2457197.9) to 3 October 2015 (BJD=2457298.7). The HIRES measurements are characterized by a RV precision of 1.5~m~s$^{-1}$ and a RMS of 5.2~m~s$^{-1}$.

As complementary data to the spectroscopic dataset, we downloaded the available photometric light curve obtained by the Kepler Space Telescope \textit{K2} mission \citep{2014PASP..126..398H} from the  Mikulski Archive for Space Telescopes (MAST). This photometric dataset was taken in the long cadence mode characterized by 30-min integration time. This time-series covers a time-span of about 78d (one Kepler quarter): from 24 August 2014 (BJD=2456894.4) to 10 November 2014 (BJD=2456972.0), which corresponds to the Campaign 2 of the \textit{K2} mission.

Finally, we included the All-Sky Automated Survey for Supernovae (ASAS-SN, \citealt{2014AAS...22323603S,2017PASP..129j4502K}) light curves of K2-38 in our photometric analysis \footnote{ASAS-SN data can be downloaded from \url{https://asas-sn.osu.edu}}. These light curves cover nearly 8 consecutive years of photometric observations in the $g$ and $V$ bands, all of which preceed the ESPRESSO RV measurements. K2-38 lies at a relatively high Galactic latitude ($b = +22$ deg) and does not have any other bright stars in its surroundings. This guarantees that ASAS-SN photometry, which is extracted from images with 8\arcsec~pixels and $\approx$15\arcsec~full width at half maximum (FWHM), is not contaminated by nearby stars. The dispersion of the $g$- and $V$-band light curves (after removing the few most deviant data points) is 33 and 62 mmag, respectively.

\section{K2-38}

\label{sec:K2-38}

\subsection{Stellar parameters}

We performed an analysis of stellar parameters using the 43 ESPRESSO spectra of K2-38 obtained during a time-span of 240 d. The blaze-corrected bi-dimensional (S2D) spectra at the barycentric reference frame were coadded, normalized, merged and corrected for RV using the StarII workflow of the data analysis software (DAS) of ESPRESSO \citep{2018SPIE10704E..0FD}. The final ESPRESSO RV-corrected normalized 1D spectrum of K2-38 (see Fig.~\ref{fig:K2-38_Spectrum}) presents a S/N of $\sim$ 230, 530, 680 and 730 at 400, 500, 600 and 700~nm, respectively, for a pixel size of 0.5~km~s$^{-1}$.

The stellar atmospheric parameters of K2-38, namely effective temperature $T_{\rm eff}$, surface gravity $\log g$, microturbulence $\xi$, and metallicity $[Fe/H]$, were derived using the {\sc ARES+MOOG} method \citep{2014dapb.book..297S}, which has been used to derive homogeneous spectroscopic parameters for the Sweet-CAT catalog \citep{2013A&A...556A.150S,2018A&A...620A..58S}. The spectral analysis is based on the excitation and ionization balance of iron. The equivalent widths (EWs) of the lines were consistently measured with the ARES (v2) code \citep{2007A&A...469..783S,2015A&A...577A..67S}, deriving the abundances in local thermodynamic equilibrium (LTE) with the spectral synthesis code MOOG (v2014) \citep{1973PhDT.......180S}. For this step, we used a grid of plane-parallel Kurucz ATLAS9 model atmospheres \citep{1993ASPC...44...87K}. The line list used for this analysis is the same as in \citet{2008A&A...487..373S}. This method provided the parameters ($T_{\rm eff}$, $\log g$, $\xi$, $[Fe/H]$)  = (5731$\pm$66~K, 4.38$\pm$0.11~dex, 0.98$\pm$0.04~km~s$^{-1}$, 0.26$\pm$0.05~dex). To cross-check these results, we used the {\sc StePar} code \citep{2019A&A...628A.131T} based on the MARCS models \citep{2008A&A...486..951G}, the MOOG code (2017 version) and the TAME code \citep{2012MNRAS.425.3162K}. Measuring the EWs of FeI-II lines \citep{2015PhyS...90e4010H}, we obtained the parameters ($T_{\rm eff}$, $\log g$, $\xi$, $[Fe/H]$) = (5754$\pm$83~K, 4.32$\pm$0.18~dex, 0.85$\pm$0.14~km~s$^{-1}$, 0.33$\pm$0.07~dex). We adopted the values from the ARES2+MOOG2014 method (which falls within the Stepar errorbars).

We used the PARAM code \citep{2006A&A...458..609D,2017MNRAS.467.1433R} to compute the mass and radius of the star through a grid-based approach, matching the stellar parameters $T_{\rm eff}$, $\log g$, and $[Fe/H]$ obtained in the previous analysis, along with the Gaia DR2 parallax and the $V$ magnitude published in the literature, to a grid of stellar evolutionary tracks and isochrones from PARSEC \citep{2012MNRAS.427..127B}. Our optimization method is based on the PARAM2 implementation \citep{2014MNRAS.445.3823R}, and provides posterior probability distribution functions (PDFs) for the stellar properties using a set of input parameters that are considered at once. The method uses the absolute magnitude computed through the model along with the apparent magnitude to derive the distance of the star and the extinction coefficient. From this method we obtained a stellar mass of $M_{\star}$~=~1.03$^{+0.04}_{-0.02}$~M$_{\odot}$ and a stellar radius of $R_{\star}$~=~1.06$^{+0.09}_{-0.06}$~R$_{\odot}$. For consistency, we also derived the mass and radius for the star using the \citet{Torres-2010} calibration. We considered the same values of $T_{\rm eff}$, $\log g$, and $[Fe/H]$ used in the previous analysis (correcting the $\log g$ value using the \citet{2014A&A...572A..95M} calibration for transits) and carried out 10000 Monte-Carlo trails. The final mass and radius values were estimated from the average of the resulting distributions, and the 1-sigma dispersion was used to estimate their uncertainties. Finally, the value for the stellar mass was corrected using the factor described in \citep{2013A&A...556A.150S}, to take into account the existing offset with respect to the isochrone masses. This approach provides a mass of $M_{\star}$~=~1.06$\pm$0.02~M$_{\odot}$ and a radius of $R_{\star}$~=~1.16$\pm$0.06~R$_{\odot}$. These values are compatible, within the errorbars, with the ones estimated using the isochrone analysis. The final adopted stellar parameters of K2-38 are listed in Table~\ref{tab:K2-38_Propiedades}.

\renewcommand{\arraystretch}{1.25}

\begin{table}
\centering
	\caption{Stellar properties of K2-38.}
	\label{tab:K2-38_Propiedades}
	\begin{tabular}{lccr} 
		\hline
		Parameter & K2-38 & Ref.\\
		\hline
		$RA$ (J2000)                               & 16:00:08.06                     & [1] \\
		$DEC$ (J2000)                              & -23:11:21.33                    & [1] \\
		$\mu_{\alpha} \cos \delta$ (mas yr$^{-1}$) & -57.00 $\pm$ 0.10               & [1] \\
		$\mu_{\delta}$ (mas yr$^{-1}$)             & -37.63 $\pm$ 0.06               & [1] \\
		Parallax (mas)                             & 5.16 $\pm$ 0.07                 & [1] \\
		Distance [pc]                              & 193.6 $\pm$ $^{+2.7}_{-2.5}$    & [2] \\
		$m_{\rm B}$                                & 12.27  $\pm$ 0.13               & [3] \\
		$m_{\rm V}$                                & 11.39 $\pm$ 0.03                & [3] \\
		Spectral type                              & G2V                             & [4] \\
		Age [Gyr]                                  & 6.7 $^{+2.4}_{-3.0}$            & [2] \\ 
		\textit{T}$_{\rm eff}$ [K]                 & 5731 $\pm$ 66                   & [2] \\
	    $[Fe/H]$ (dex)                             & 0.26 $\pm$ 0.05                 & [2] \\
		$M_{\star}$ [M$_{\odot}$]                  & 1.03 $^{+0.04}_{-0.02}$         & [2] \\
		$R_{\star}$ [R$_{\odot}$]                  & 1.06 $^{+0.09}_{-0.06}$         & [2] \\
		$L_{\star}$ [L$_{\odot}$]                  & 1.09 $\pm$ 0.15                 & [1] \\
		$\log g$ (cgs)                             & 4.38 $\pm$ 0.11                 & [2] \\
		$\xi$ [km s$^{-1}$]                        & 0.98 $\pm$ 0.04                 & [2] \\
		$A_{V}$                                    & 0.15 $^{+0.16}_{-0.14}$         & [2] \\
		$v \sin i$ [km s$^{-1}$]                   & $<$ 2                           & [4] \\  
		$\log_{10}$ ($R^{'}_{\rm HK}$)             & -5.06 $\pm$ 0.13                & [2] \\
		\hline
	\end{tabular}
	\begin{minipage}{\columnwidth} 
{\footnotesize \textbf{References:} [1] \citet{2018A&A...616A...1G}; [2] This work; \newline [3] \citet{2016yCat.2336....0H}; [4] \citet{2016ApJ...827...78S}};
\end{minipage}	
\end{table}

\subsection{Chemical abundances}

Using the adopted stellar parameters we measured the element abundances of C, O, Mg, and Si using the aforementioned tools (MOOG and ARES) by closely following the methodology described in previous works by \citet{2015A&A...583A..94A,2016A&A...591A..34A}. We calculated the chemical abundances of these elements from the EWs of each spectral line, using the LTE code MOOG \citep{1973PhDT.......180S} with an appropriate ATLAS model atmosphere \citep{1993ASPC...44...87K} of K2-38. The final abundances were computed as the average value of individual element abundances. The errors were estimated from the sensitivities of the element abundances to the uncertainties on the stellar parameters added quadratically to the dispersion from the individual element abundances. We derived the following abundances: $[C/H]$~=~0.21~$\pm$~0.06, $[O/H]$~=~0.18~$\pm$~0.07, $[Mg/H]$~=~0.24~$\pm$~0.05, $[Si/H]$~=~0.27~$\pm$~0.06. The atomic line parameters including oscillator strengths of spectral lines of MgI and SiI were taken from \citet{2012A&A...545A..32A} (two Si lines at $\lambda$=5701.11~\AA \ and $\lambda$=6244.48~\AA \ were excluded following the recomendation of \citealt{2016A&A...591A..34A}). The oxygen abundances were determined using two weak lines at 6158.2~\AA \ and 6300.3~\AA \ following the work of \citet{2015A&A...576A..89B}. Carbon abundances were based on the two well-known CI optical lines at 5052~\AA \ and 5380~\AA. The atomic data of these lines were extracted from VALD3  database \footnote{source: \url{http://vald.astro.univie.ac.at/~vald3/php/vald.php?newsitem=0}}.

\citet{2015A&A...577A..83D} proposed that Mg/Si and Fe/Si mineralogical ratios can be used as probes to constrain the internal structure of terrestrial planets. These models were successfully tested on three terrestrial planets by \citet{2015A&A...580L..13S}. The model of \citet{2015A&A...580L..13S} was then used to explore the possible compositions of planet-building blocks and planets orbiting stars belonging to different Galactic populations \citep{2017A&A...608A..94S}. We transformed the chemical abundances of C, O, Mg, Si, and Fe relative to the Sun to absolute abundances accepting the solar reference abundances as given in \citet{2009ARA&A..47..481A} for Fe ($\log \epsilon$ = 7.5 dex), Mg ($\log \epsilon$ = 7.60 dex), and Si ($\log \epsilon$ =7.51 dex), and as given by \citet{2015A&A...576A..89B} and \citet{2017A&A...599A..96S} for O ($\log \epsilon$=8.71 dex) and C ($\log \epsilon$=8.50 dex). By using these absolute abundances we applied the aforementioned stoichiometric model of \citet{2015A&A...580L..13S} to determine the iron-mass fraction and water-mass fraction of the planet building blocks in the planetary disks of the star. Our model suggests an iron-mass fraction of 33.4 $\pm$ 3.3 \% and water-mass fraction of 51.9 $\pm$ 5.9 \%. We note that this model predicts an iron-mass fraction of 33\% and water-mass fraction of 60\% for the solar system planet building blocks \citep{2017A&A...608A..94S}. These results may have implications on the possible bulk composition of the planets formed in the planetary system of K2-38 (see Sect.~\ref{sec:Discussion}), although they are not directly applicable to the final composition or internal structure of a final differentiated planet. The final composition and structure of differentiated planets will depend on several factors such as the position in the proto-planetary disk where the planet is formed and its migration path. Additionally, some physical processes, such as evaporation and collisional stripping, may change the overall composition of planets. 

\section{Data analysis}

\label{sec:Analysis}

\subsection{Photometric analysis}

For the photometric analysis, we used the \texttt{everest} code to detrend the K2 light curve \citep{2016AJ....152..100L,2018AJ....156...99L}, which has been proved to provide better results in terms of flux scatter in comparison with other pipelines \citep{2018AJ....155..127H}. This code performs a flux correction based on a single cotrending basis vector (CBV). We performed a sigma-clipping procedure to these flux values using a 2.5$\sigma$ in order to remove outliers. We also discarded the photometry taken between BJD=2456926.85 and BJD=2456930.62 due to a flare event. Finally, we used a moving-average to clean the light curve from the small-scale outliers. 

\begin{figure} 
	\includegraphics[width=\hsize]{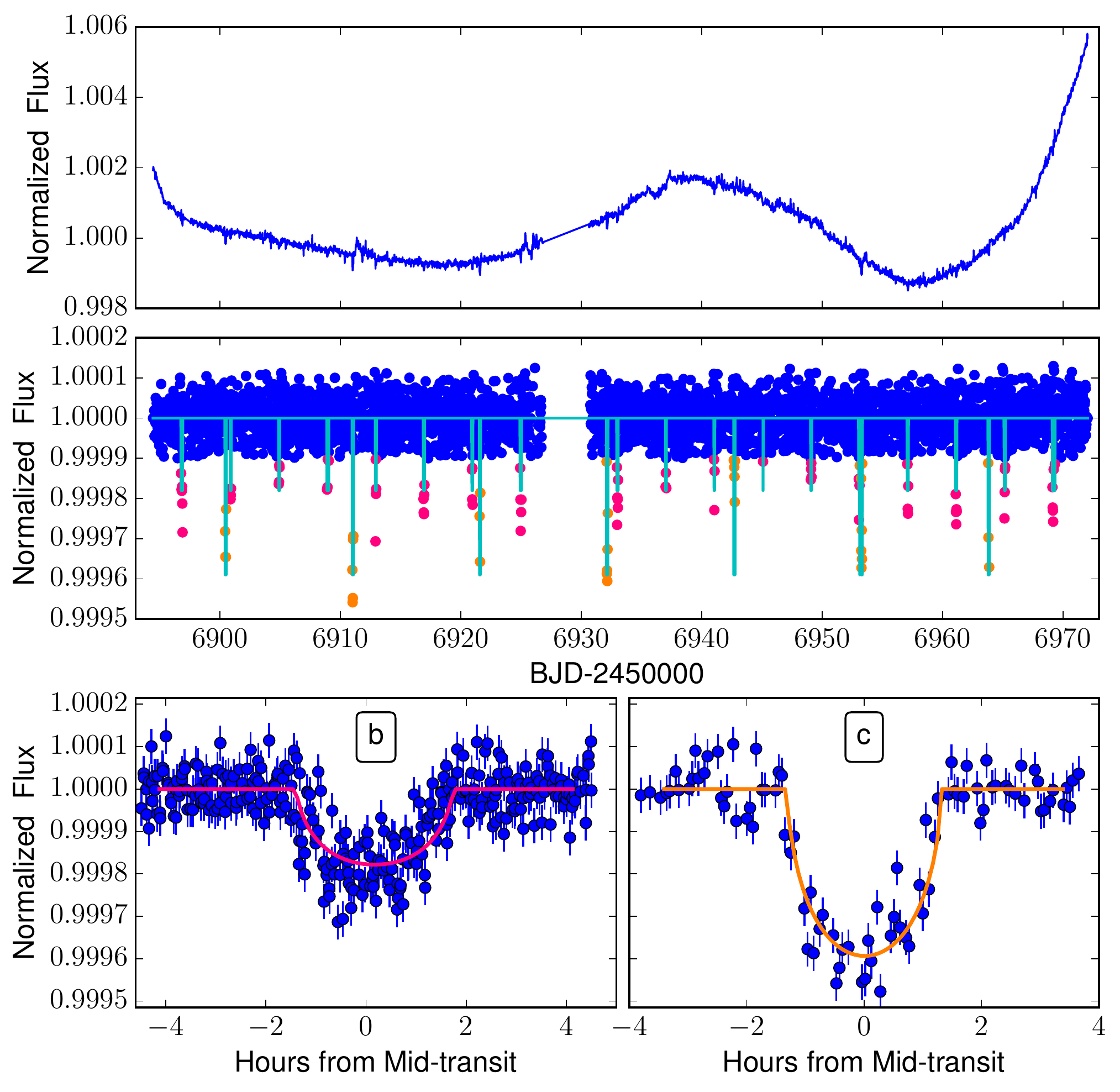} 
    \caption{\textbf{Top:} Normalized flux of the light curve extracted using the EVEREST pipeline. \textbf{Middle:} Detrended light curve with the transits of K2-38b and K2-38c marked with different colors. \textbf{Bottom:} Folded light curve using the period of K2-38b (P=4.02d) and K2-38c (P=10.56d).}
    \label{fig:Transits_K2-38_Kepler}
\end{figure}

Then we carried out a transit analysis using the \texttt{exotrending} code \citep{2017ascl.soft06001B}. We used the period and epoch of transit provided by \citet{2016ApJ...827...78S} for both known planets as input parameters and applied the model from \citet{2002ApJ...580L.171M}. Assuming a quadratic law for the limb-darkening \citep{2000A&A...363.1081C} based on two coefficients $\gamma_{1}$ and $\gamma_{2}$, we obtained the results shown in Fig.~\ref{fig:Transits_K2-38_Kepler}. The light curve shown in the top panel of Fig.~\ref{fig:Transits_K2-38_Kepler} exhibits a low-frequency modulation which is connected with the CBV correction made by \texttt{everest} and not related to the stellar activity of the star since it is not present in the light curve obtained with other detrending tools. The brightness dips shown in the bottom panel of Fig.~\ref{fig:Transits_K2-38_Kepler} are related to a $R_{\rm p}$=1.54$\pm$0.14~R$_{\oplus}$ for planet b and $R_{\rm p}$=2.29$\pm$0.26~R$_{\oplus}$ for planet c. These values are in good agreement with those reported in \citet{2016ApJ...827...78S}, with the difference between both studies coming from the improved star radius used in our work.

\begin{table}
\centering
	\caption{Planetary parameters of K2-38b and K2-38c obtained in the photometric analysis.}
	\label{tab:K2-38_PhotometricProperties}
	\begin{tabular}{lcr} 
		\hline
		Parameter                      &  K2-38b                     & K2-38c                  \\
		\hline
		$P$ [days] $^{(*)}$            & 4.01593$\pm$0.00050              & 10.56103$\pm$0.00090        \\
		$T_{0}$ [BJD-2456000] $^{(*)}$ & 896.8786$\pm$0.0054              & 900.4752$\pm$0.0033         \\
		$a/R_{\rm \star}$              & 10.13$^{+0.75}_{-0.66}$          & 19.30$^{+1.43}_{-1.26}$     \\
		$R_{\rm p}$/$R_{\rm \star}$    & 0.0133$^{+0.00071}_{-0.00076}$   & 0.020$^{+0.0017}_{-0.0018}$ \\
		$\boldsymbol{\rm \gamma_{1}}$  & 0.4770$\pm$0.0060                & 0.4770$\pm$0.0060           \\
		$\boldsymbol{\rm \gamma_{2}}$  & 0.2040$\pm$0.0070                & 0.2040$\pm$0.0070           \\
		\hline
	\end{tabular}
	\begin{minipage}{\columnwidth} 
	{\footnotesize $^{(*)}$ \citet{2016ApJ...827...78S}.}
\end{minipage}	
\end{table}

The GLS periodograms of the ASAS-SN light curves show a long-term trend ($>$1000-2000 d). When this trend is removed with a linear fit, the remaining periodogram shows no significant peaks above the 10\%~False Alarm Probability (FAP) level. We also obtained the GLS periodogram of the merged $g$- and $V$-band light curves, which may improve the S/N detection of any periodic signal, with a similar result: no strong peak between 2 and 1000 days. We concluded that ASAS-SN data confirm that K2-38 is photometrically inactive above ~30–60 mmag (1 $\sigma$) for nearly 8 years of observations before ESPRESSO.

\subsection{Spectroscopic analysis}

\begin{figure} 
	\includegraphics[width=\hsize]{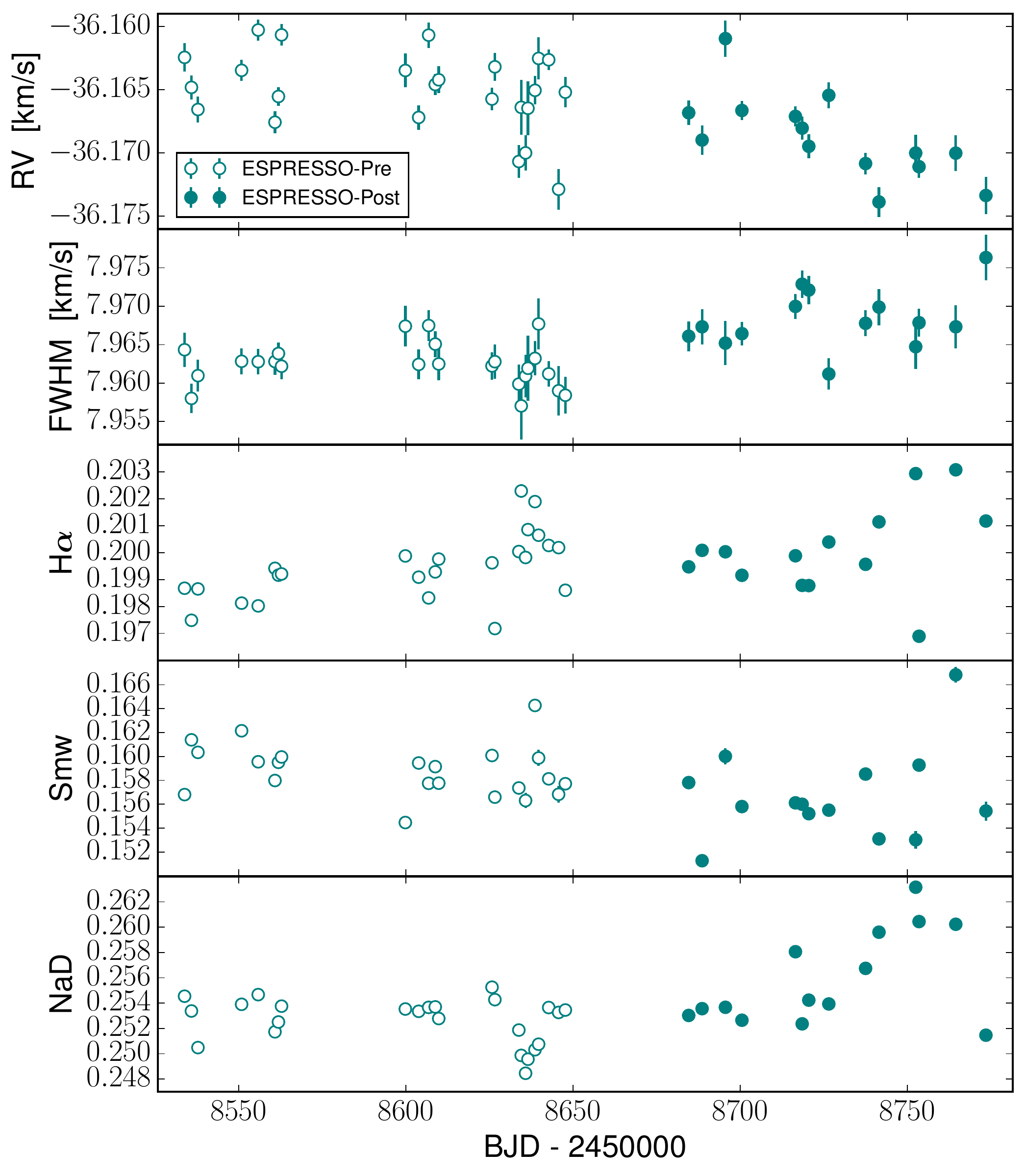} 
    \caption{Time-series of the RVs and stellar indices obtained with the ESPRESSO pipeline: RV, FWHM, H$_{\alpha}$, S-index, and NaD index.}
    \label{fig:Compact_Indexes_Spec+Phot_K2-38}
\end{figure}

For the spectroscopic analysis, we recovered the RVs computed with the ESPRESSO pipeline using the DACE interface \footnote{The Data Analysis Center for Exoplanets (DACE) platform is available at \url{https://dace.unige.ch}}, along with different stellar activity indicators associated with certain spectral lines: the H$_{\alpha}$ index \citep{2011A&A...534A..30G}, the S$_{\rm MW}$ index related to the CaII H\&K lines \citep{2011arXiv1107.5325L}, and the NaD index related to the NaI D$_{1}$ and D$_{2}$ lines \citep{2007MNRAS.378.1007D}. We also measured the FWHM of the CCF. The values obtained are shown in Fig.~\ref{fig:Compact_Indexes_Spec+Phot_K2-38}.

\subsubsection{Stellar activity}

Using the S$_{\rm MW}$ time-series we calculated a chromospheric activity level \citep{1984ApJ...279..763N} of $\log (R^{'}_{\rm HK}$)~=~-5.06~$\pm$~0.13 following the procedure described in \citet{2015MNRAS.452.2745S}. Using the chromospheric activity-rotation relation obtained by \citet{2016A&A...595A..12S} we estimated an expected rotation period of $P_{\rm rot}$~=~29~$\pm$~2~d.

We performed a stellar activity analysis using the different spectroscopic indices available to search for the rotation signal. To study the main periodicities present in the time-series, we built a Generalized Lomb-Scargle (GLS) periodogram \citep{2009A&A...496..577Z} for each one of them. To establish how reliable are the signals we were detecting, we calculated the theoretical 10, 1 and 0.1\%~FAP levels through a bootstrapping process \citep{2004MNRAS.354.1165C}.

All the signals in the H$_{\rm \alpha}$, S$_{\rm MW}$, NaD, and FWHM periodograms had a significance below the 10\% level, hence no significant periodicity was detected in any of the datasets. This points out to the fact that the rotation signal may have an amplitude lower than the RMS of the data. The time-span of observations does not cover the typical timescale of signals associated with long-term magnetic cycles, and therefore, we cannot assess their presence on this star with the current dataset. We performed a Gaussian Processes (GP) analysis in the four time-series separately based on the \texttt{celerite} code \citep{2017AJ....154..220F}. For modeling the rotation we built the kernel described in the same article:

\begin{equation}
  \kappa(\tau)=\frac{A}{2+C}e^{-\tau/t_{s}}\left[\cos\left(\frac{2 \pi \tau}{P_{rot}}\right)+(1+C)\right]
  \label{eq:Kernel}
\end{equation}

\noindent where \textit{A} is the amplitude of the covariance, $t_{\rm s}$ the coherence time of the signal, $P_{rot}$ the rotation period, and \textit{C} the scale parameter that weights the periodic and nonperiodic components of the model. We introduced an additional jitter term for each dataset, which leads to the built of two kernels (one for each dataset) with the same rotation parameters. We set up our GP model as a combination of these two kernels with boundaries for all the parameters. We used the expected rotation value derived from the log(R$^{'}_{\rm HK}$) index as the first guess for $P_{\rm rot}$, and the double of this value as the first guess for $t_{\rm s}$ \citep{2017MNRAS.472.1618G}. We then defined our likelihood function including two offset terms to account for the zero-point of each dataset. We minimized this function using the \texttt{minimize} python module included in the \texttt{scipy.optimize} package. The parameters obtained from this analysis performed on the four time-series did not provide any clear rotation value for the star. We also carried out a MCMC analysis on these time-series using the \texttt{emcee} python code \citep{2010CAMCS...5...65G,2013PASP..125..306F}. We established priors based on the results from the previous \texttt{minimize} analysis, but the rotation parameters did not converge to a clear value. We repeated this GP analysis on the RV time-series and we did not find any conclusive result regarding the activity of the star.

\subsubsection{Planetary characterization}

Due to the lack of a good characterization of the rotation of the star, we opted for simplifying our activity model, maintaining only the jitter terms. We performed two separate analyses on the RV time-series using the \texttt{minimize} and \texttt{emcee} packages in each one of them. We repeated the two-step structure carried out in the stellar indices time-series, using first the \texttt{minimize} package to obtain preliminary results that are considered to establish the priors for the \texttt{emcee} package. For the MCMC analysis, we defined a sample of 512 walkers, running first a burn-in chain with 10\,000 steps, followed by a construction chain with 50\,000 steps. All the figures shown in the rest of this Section were obtained using the MCMC results.

\begin{figure}  
	\includegraphics[width=\hsize]{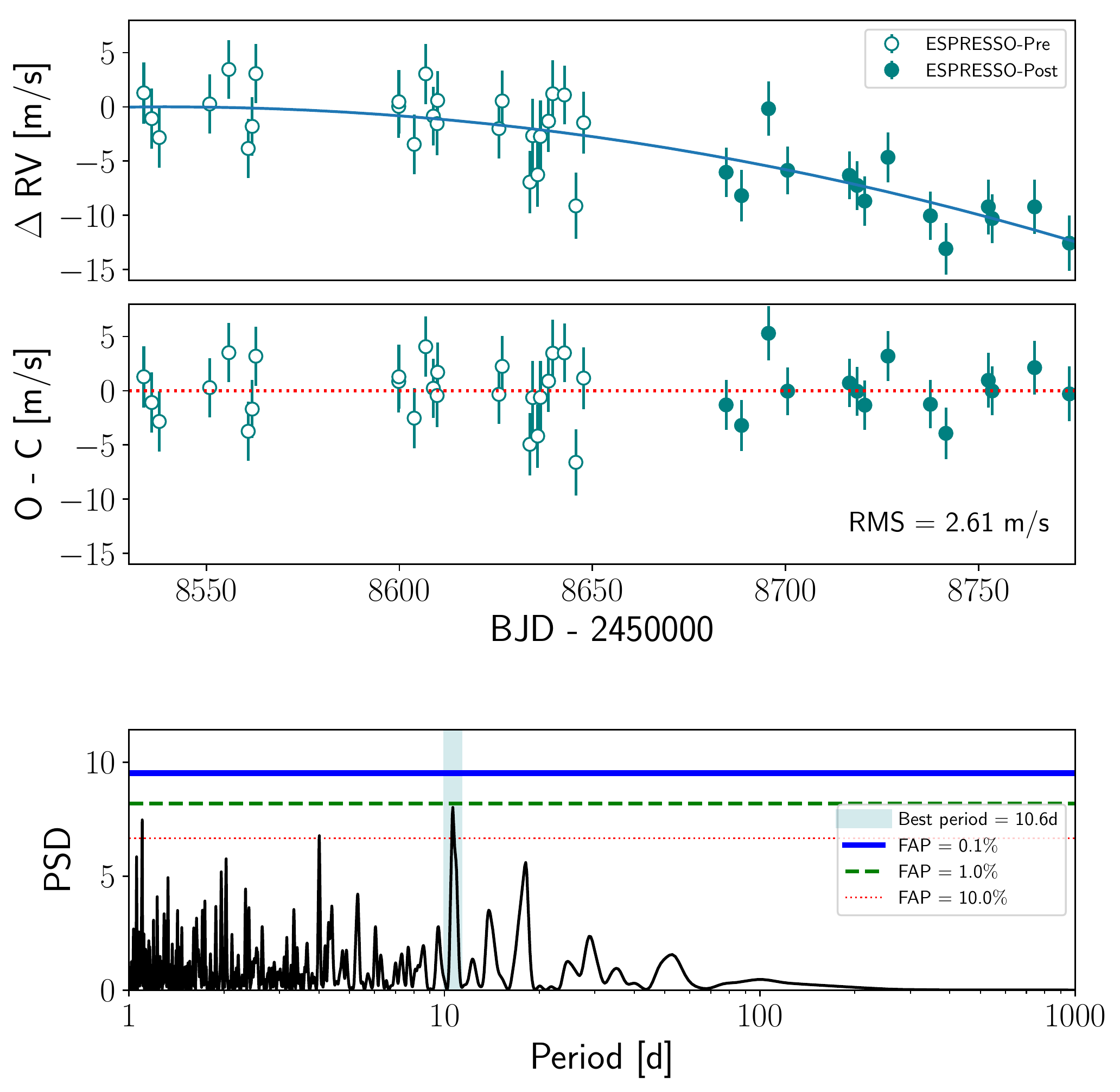}
    \caption{\textbf{Top:} ESPRESSO RV time-series with the trend model. \textbf{Center:} Residuals after subtracting the model. \textbf{Bottom:} Periodogram of the residuals.}
    \label{fig:3_2_RV}
\end{figure}

We first modeled the RV time-series with a quadratic polynomial to account for a visible trend in the time-series, along with two offset terms to account for the zero-point of each dataset. The minimization of all the parameters provided the results shown in Fig.~\ref{fig:3_2_RV}.

\begin{figure} 
	\includegraphics[width=\hsize]{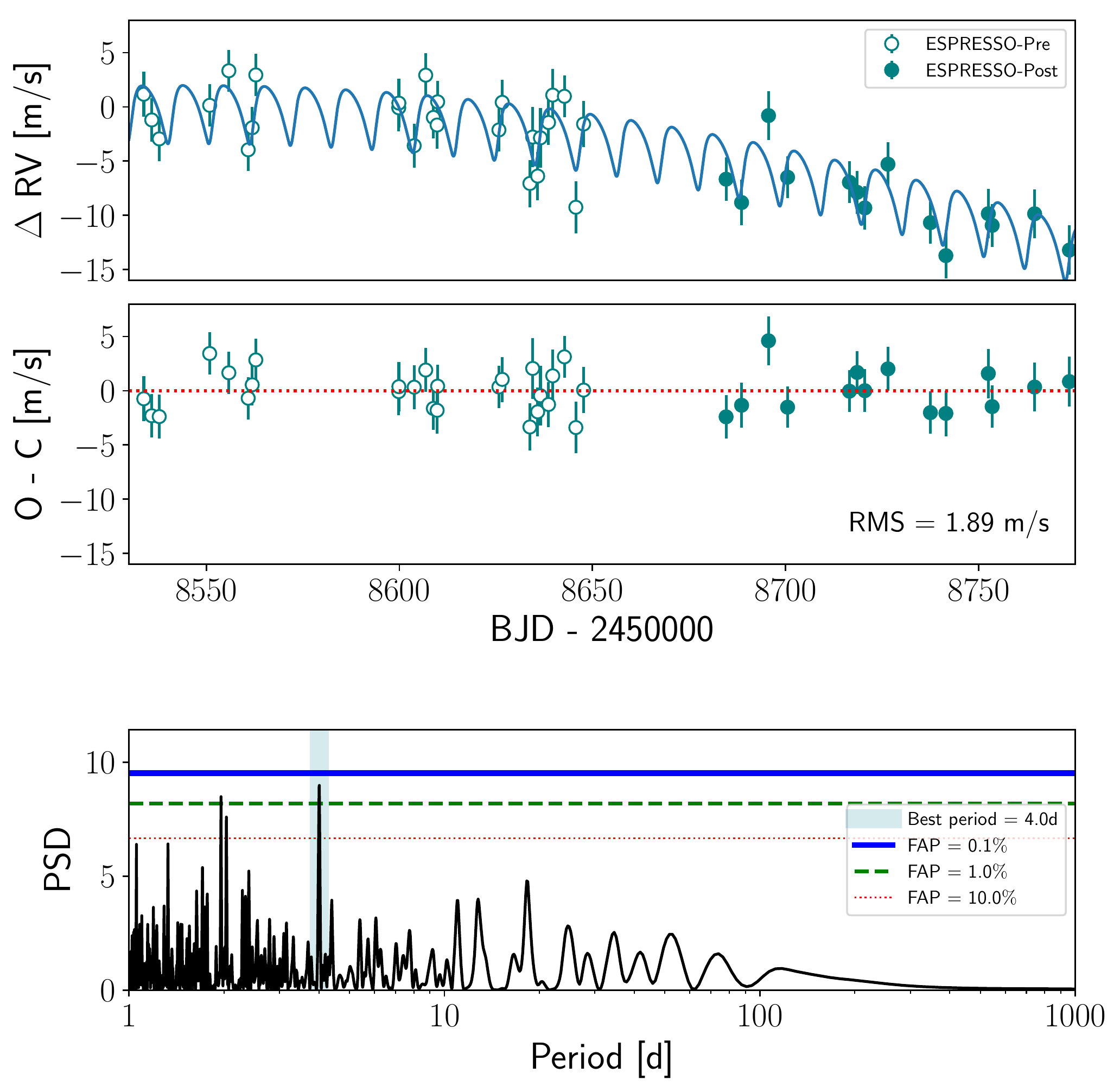} 
    \caption{\textbf{Top:} ESPRESSO RV time-series with a model that includes a trend and the planetary signal from K2-38c. \textbf{Center:} Residuals after subtracting the model. \textbf{Bottom:} Periodogram of the residuals.}
    \label{fig:3_5_RV}
\end{figure}

This first model leaves a RMS in the residuals of 2.61~m~s$^{-1}$. In the periodogram of Fig.~\ref{fig:3_2_RV} the signal of the planet c appears with a FAP close to 1\%. We modeled this signal in the original RV time-series with the following Keplerian \citep{2016A&A...590A.134D}:

\begin{equation}
  y(t)=K \left(\cos(\nu+\omega) + e \ \cos(\omega)\right)
  \label{eq:Keplerian}
\end{equation}

\noindent where the true anomally $\nu$ is related to the solution of the Kepler equation that depends on the orbital period of the planet P$_{\rm orb}$ (obtained by photometric analysis) and the orbital phase $\phi$. This phase corresponds to the periastron time, which depends on the mid-point transit time $T_{0}$, the argument of periastron $\omega$, and the eccentricity of the orbit $e$. We used the $T_{0}$ from the photometric analysis and started assuming a circular orbit (i.e., $\rm \omega$=0 and $e$=0) for the first guess of the periastron time. Therefore, we added only three parameters to our model (the amplitude of the signal $K$, $\omega$, and $e$). We then carried out the minimization of all the parameters (recomputing the ones related to the trend, offsets, and jitter) using the original RV time-series. This provided the results shown in Fig.~\ref{fig:3_5_RV}.

\begin{figure}
	\includegraphics[width=\hsize,page=1]{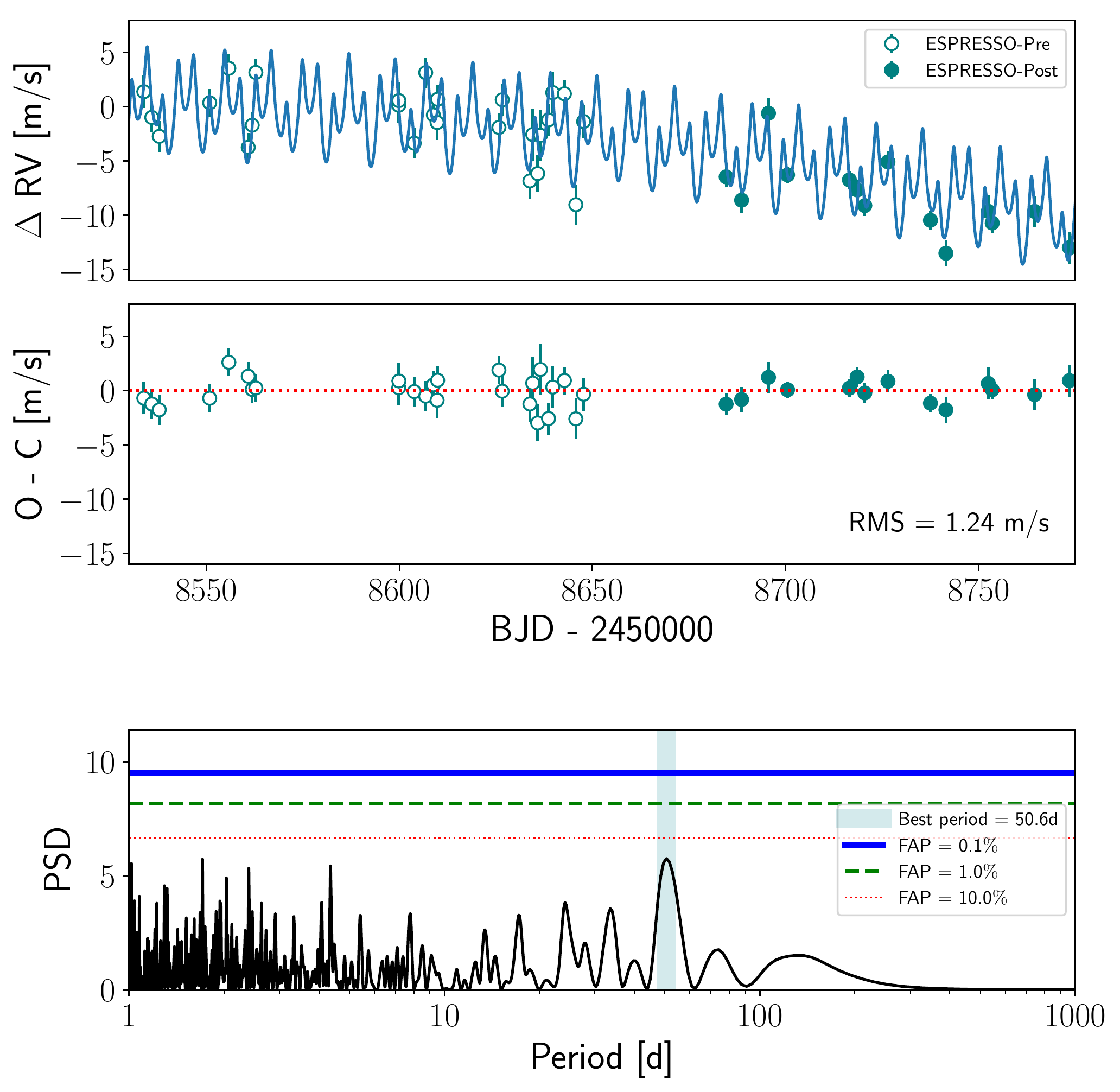}  
    \caption{\textbf{Top:} ESPRESSO RV time-series with a model that includes a trend and the planetary signals from K2-38b and K2-38c. \textbf{Center:} Residuals after subtracting the model. \textbf{Bottom:} Periodogram of the residuals.}
    \label{fig:RV_act_2pla}
\end{figure}

This model lowers the RMS of the residuals to 1.89~m~s$^{-1}$. In the bottom panel of Fig.~\ref{fig:3_5_RV} the signal from K2-38b appears with a FAP below the 1\%. We included this signal in our model as a second Keplerian using the period shown in Table~\ref{tab:K2-38_PhotometricProperties}, which adds four new parameters to fit. From the minimization of the 12 parameters based on the original RV time-series, we obtained the results shown in Fig.~\ref{fig:RV_act_2pla}.

\begin{figure}  
	\includegraphics[width=\hsize]{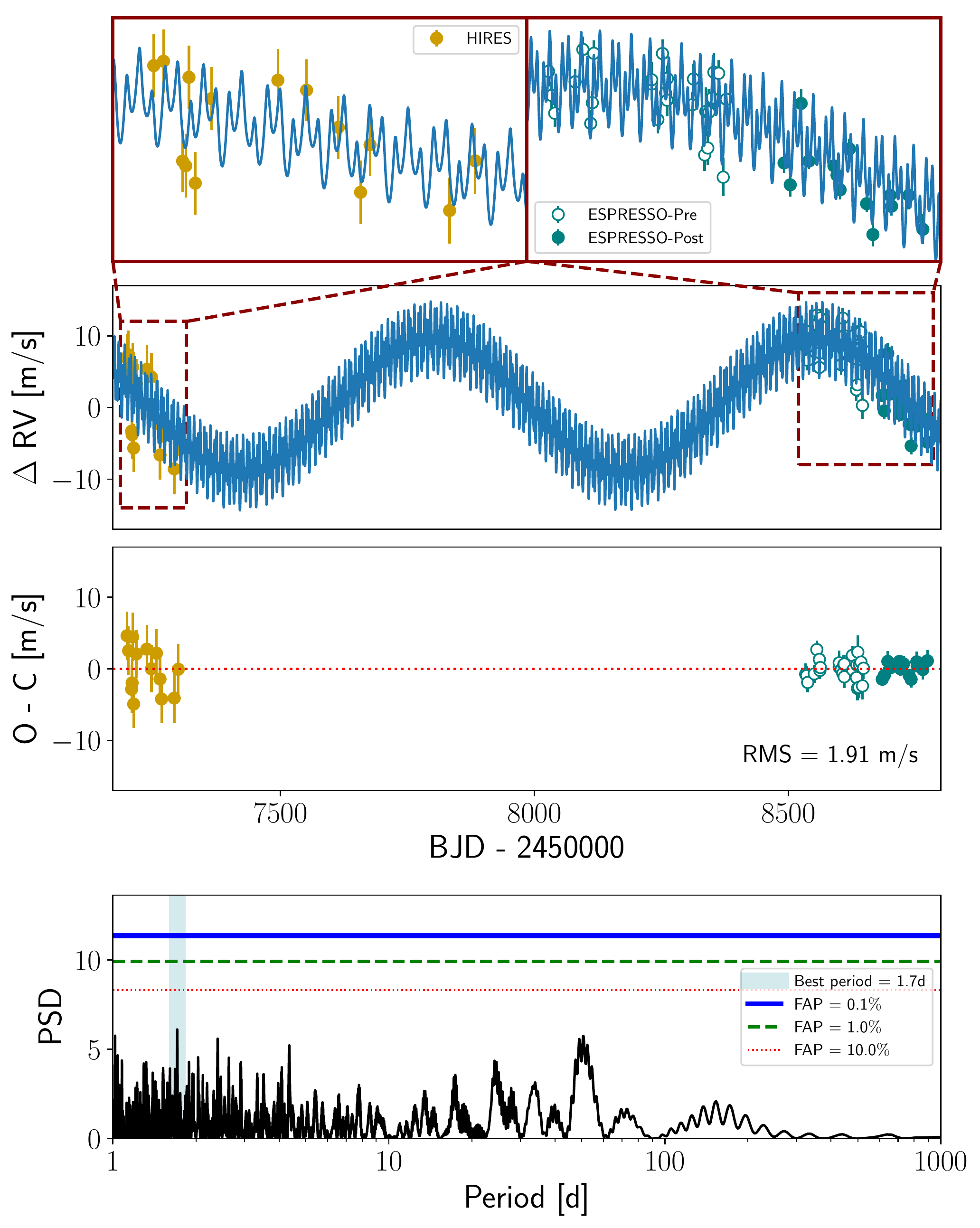}
    \caption{\textbf{Top:} HIRES and ESPRESSO RV time-series with a model that includes the long-period signal fitted by a sinusoidal along with the two planetary signals fitted by Keplerians. \textbf{Center:} Residuals after subtracting the model. \textbf{Bottom:} Periodogram of the residuals.}
    \label{fig:9_12_RV}
\end{figure}

The RMS of residuals after subtracting this last model is 1.24~m~s$^{-1}$, which is very close to the time-series mean photon noise (i.e., mean of internal errors) of 1.14~m~s$^{-1}$. In the periodogram of the residuals from Fig.~\ref{fig:RV_act_2pla} is clearly seen that there are no more significant peaks. We then incorporated the HIRES dataset into the time-series with its own offset and jitter terms. In this combined dataset we treated the long-period signal with a sinusoidal since this approach provided a lower RMS in the residuals and better Bayesian evidence $\log Z$ \citep{2013arXiv1311.0674P}. The minimization of the 15 final parameters led to the results shown in Fig.~\ref{fig:9_12_RV}.

\begin{figure}
	\includegraphics[width=\hsize,page=2]{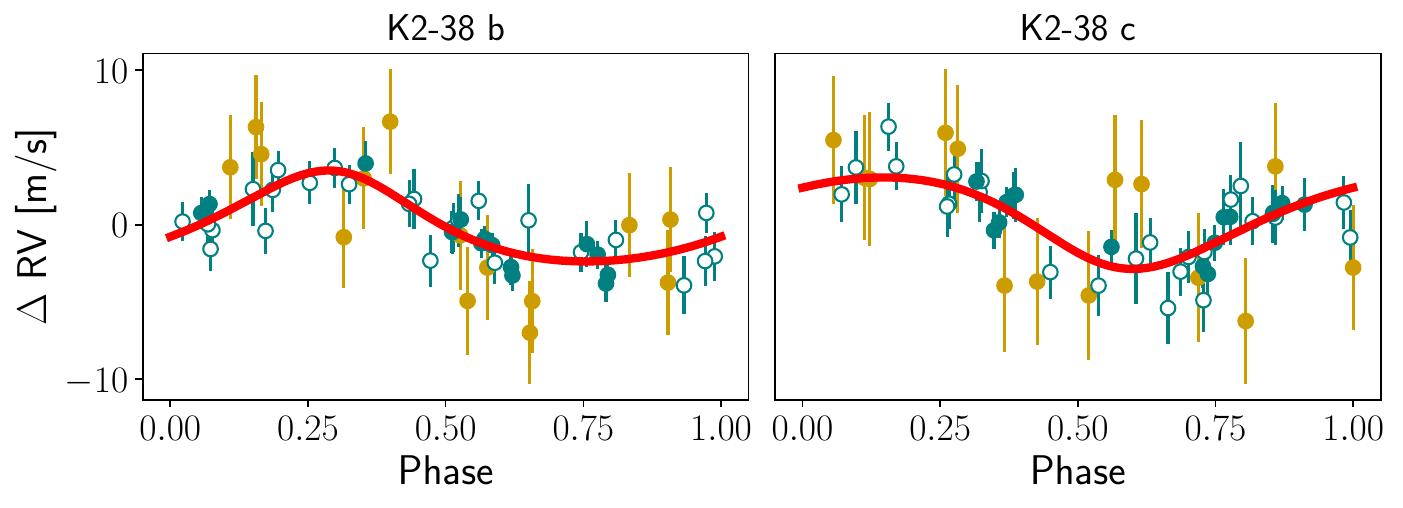} 
    \caption{Phase folded RV curve of both planets using their respective periastron time as a time reference. Open and solid blue symbols correspond to ESPRESSO data before and after the technical intervention of June 2019, respectively. Solid yellow symbols correspond to HIRES data.}
    \label{fig:RV_phaseFolded}
\end{figure}

This combined treatment provides a higher RMS in the residuals due to the lower quality of the HIRES measurements but puts better constraints of the long-period signal. We show the individual phase-folded RV models of the two planets in Fig.~\ref{fig:RV_phaseFolded}.

\begin{figure}
	\includegraphics[width=\hsize,page=2]{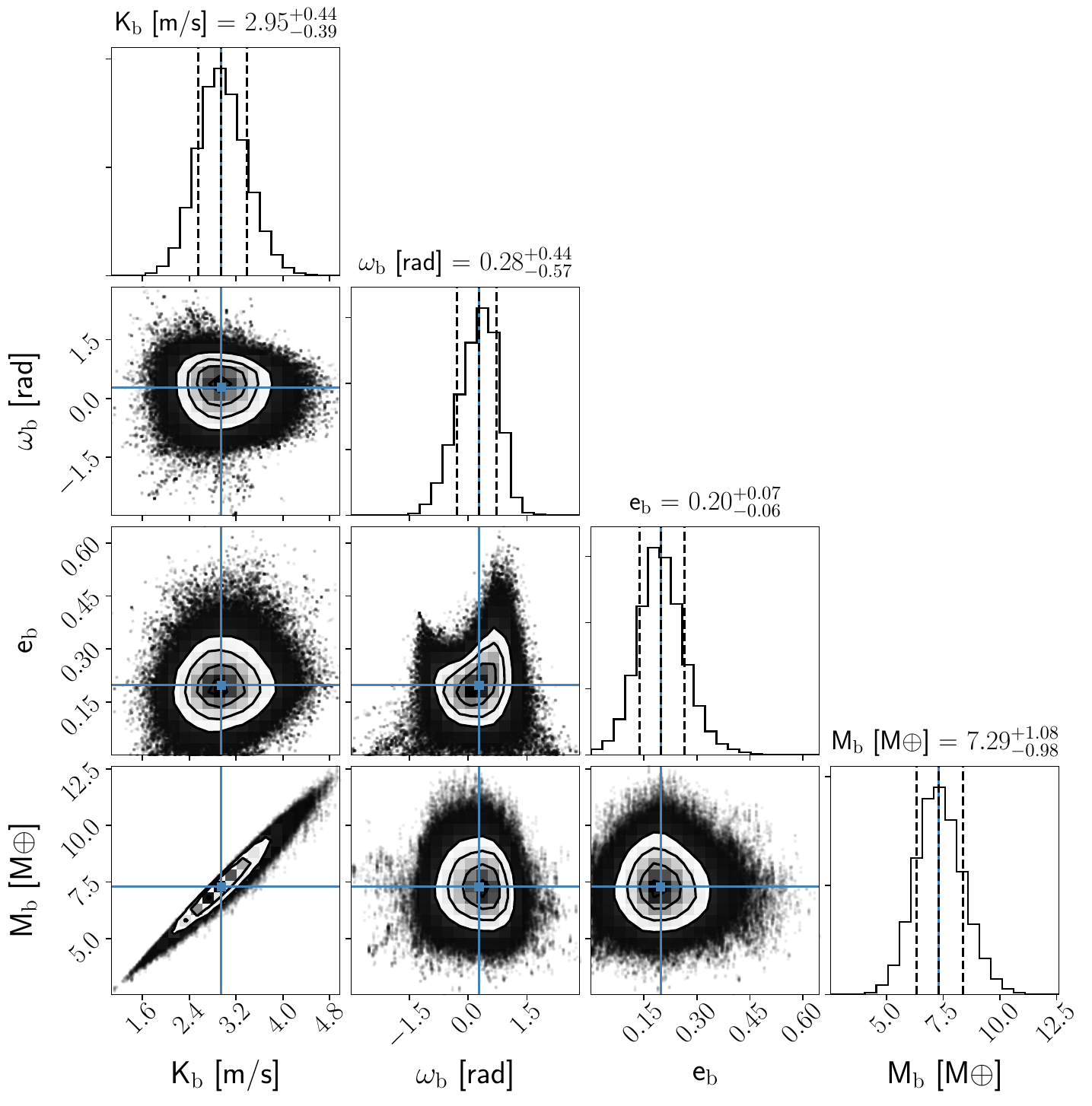} 
    \caption{Corner plot of the fitted parameters associated with K2-38c in the 2 planets model for the K2-38 RV time-series. The vertical lines indicate the mean and the 16th-84th percentiles.}
    \label{fig:RV_MCMC_c}
\end{figure}

\begin{figure}
	\includegraphics[width=\hsize,page=3]{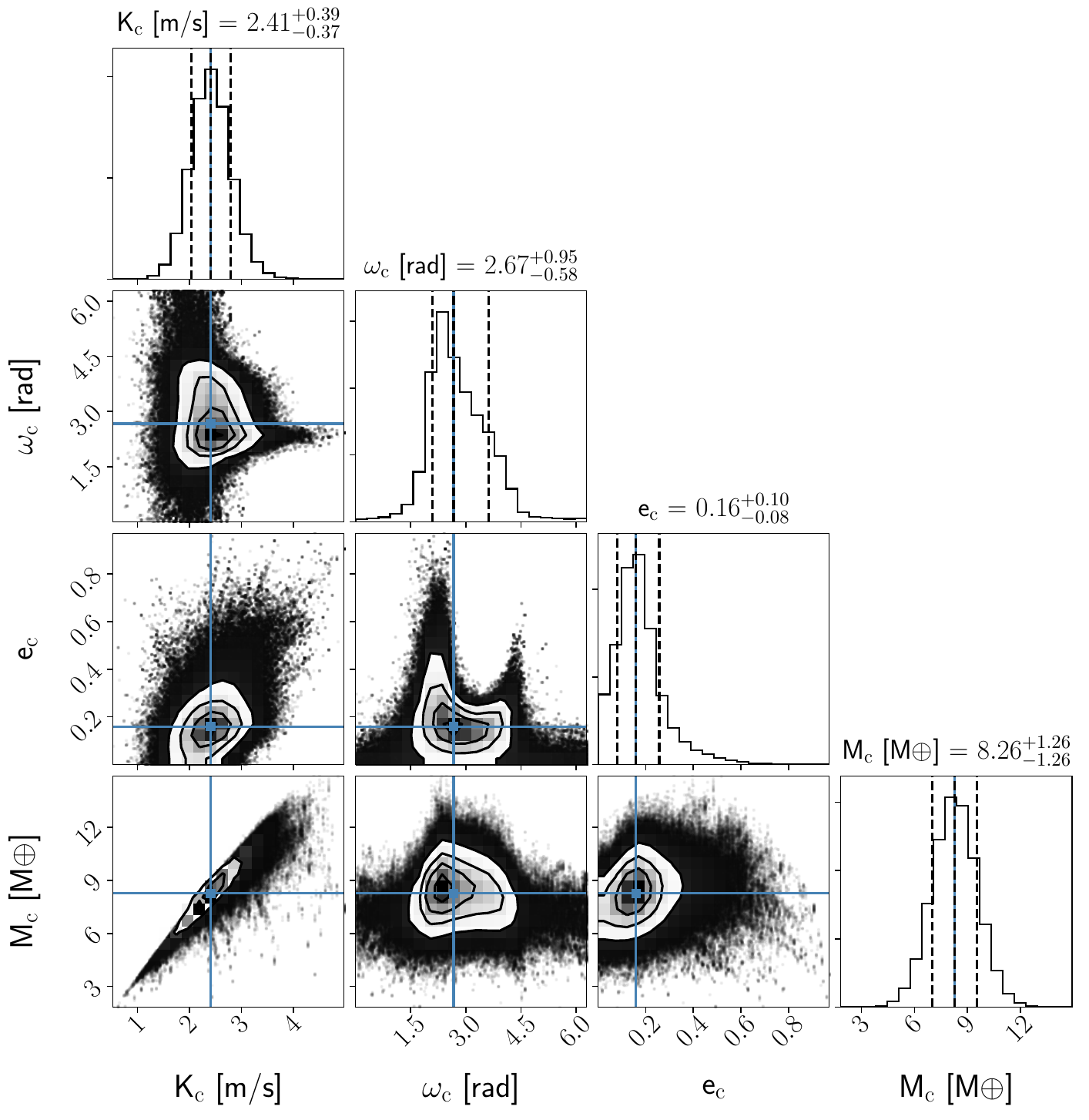} 
    \caption{Corner plot of the fitted parameters associated with K2-38b in the 2 planets model for the K2-38 RV time-series. The vertical lines indicate the mean and the 16th-84th percentiles.}
    \label{fig:RV_MCMC_b}
\end{figure}

We show all the parameters from the 2 Keplerians+trend+jitter+offset model along with the parameter priors in Table~\ref{tab:K2-38_PlanetProperties}. The parameter distribution obtained from the MCMC is shown in Fig.~\ref{fig:RV_MCMC_c} and Fig.~\ref{fig:RV_MCMC_b} for planets K2-38b and K2-38c respectively, including their significant nonzero eccentricities. The complete set of parameters is depicted in Fig.~\ref{fig:RV_MCMC}. We have included in these figures the mass of both planets calculated as:

\begin{equation}
  M_{\rm P}=K M_{\rm \star}^{2/3} P^{1/3} \frac{\sqrt{1-e^{2}}}{\sin i} \frac{317.8}{28.4329} M_{\rm \oplus}
  \label{eq:Mass}
\end{equation}

\noindent where \textit{K} is the amplitude of the planetary signal in m~s$^{-1}$, $M_{\rm \star}$ is the mass of the star in solar masses from Table~\ref{tab:K2-38_Propiedades}, \textit{P} is the orbital period of the planet in yr, \textit{e} is the eccentricity, and $i$ is the orbital inclination (we computed this last parameter from the radius of the star, duration of the transit, semi-major axis and orbital period of the planets). We obtained a value of $M_{\rm p}$=7.3$^{+1.1}_{-1.0}$~M$_{\oplus}$ for K2-38b and $M_{\rm p}$=8.3$^{+1.3}_{-1.3}$~M$_{\oplus}$ for K2-38c from the MCMC analysis. We also repeated the analysis establishing the orbital period of both planets as free parameters and recovered values of $P_{\rm orb}$=4.01$\pm$0.01~d and $P_{\rm orb}$=10.56$\pm$0.04~d for K2-38b and K2-38c, respectively, that match the results from photometry. Table~\ref{tab:K2-38_DerivedProperties} contains all the derived parameters from the MCMC analysis along with the ones from the K2 light curve analysis. 

\renewcommand{\arraystretch}{1.5}

\begin{table}
\centering
	\caption{Planetary parameters of K2-38b and K2-38c from the MCMC analysis.}
	\label{tab:K2-38_PlanetProperties}
	\begin{tabular}{lcr} 
		\hline
		Parameter & MCMC Priors & Results \\
		\hline
				&  \multicolumn{2}{c}{K2-38b} \\
		\hline
		$K$ [m s$^{-1}$]           & $\mathcal{U}$ (0.0, 5.0)         & 2.95$^{+0.44}_{-0.39}$    \\ 
		$\rm \omega$ [rad]         & $\mathcal{U}$ (-$\pi$, $\pi$)    & 0.28$^{+0.44}_{-0.57}$    \\
		$e$                        & $\mathcal{U}$ (0.0, 1.0)         & 0.197$^{+0.067}_{-0.060}$ \\
		\hline
		 & \multicolumn{2}{c}{K2-38c} \\
		\hline
		$K$ [m s$^{-1}$]           & $\mathcal{U}$ (0.0, 5.0)         & 2.41$^{+0.39}_{-0.37}$    \\
		$\rm \omega$ [rad]         & $\mathcal{U}$ (0.0, 2$\pi$)      & 2.67$^{+0.95}_{-0.58}$    \\
		$e$                        & $\mathcal{U}$ (0.0, 1.0)         & 0.161$^{+0.096}_{-0.078}$ \\
		\hline
		& \multicolumn{2}{c}{Long-Period signal} \\
		\hline		
		$K$  [m s$^{-1}$]  & $\mathcal{U}$ (0.0, 20.0)      & 9.3$^{+1.7}_{-1.6}$ \\
		$P$  [days]        & $\mathcal{U}$ (650.0, 900.0)   & 753$^{+36}_{-34}$   \\
		$T$  [BJD-2457000] & $\mathcal{U}$ (1050.0, 2050.0) & 1168$^{+15}_{-17}$  \\
		\hline
		& \multicolumn{2}{c}{Other terms} \\
		\hline
		jitter$_{\rm HIRES}$  [m s$^{-1}$] & $\mathcal{LU}$ (0.01, 10.0) & 3.02$^{+0.93}_{-0.70}$ \\
		jitter$_{\rm E-Pre}$  [m s$^{-1}$] & $\mathcal{LU}$ (0.01, 5.0)  & 0.96$^{+0.42}_{-0.51}$ \\
		jitter$_{\rm E-Post}$ [m s$^{-1}$] & $\mathcal{LU}$ (0.01, 5.0)  & 0.09$^{+0.34}_{-0.07}$ \\
		offset$_{\rm HIRES}$  [m s$^{-1}$] & $\mathcal{U}$ (-20.0, 20.0) & -0.4$^{+4.5}_{-4.1}$   \\
		offset$_{\rm E-Pre}$  [m s$^{-1}$] & $\mathcal{U}$ (-20.0, 20.0) & -7.1$^{+1.7}_{-1.8}$   \\
		offset$_{\rm E-Post}$ [m s$^{-1}$] & $\mathcal{U}$ (-20.0, 20.0) & -2.5$^{+1.5}_{-1.6}$   \\
		\hline
	\end{tabular}
	\begin{minipage}{\columnwidth} 
\end{minipage}	
\end{table}

\begin{table}
\centering
	\caption{Derived planetary parameters of K2-38b and K2-38c.}
	\label{tab:K2-38_DerivedProperties}
	\begin{tabular}{lcr} 
		\hline
		Parameter                      &  K2-38b                         & K2-38c                            \\
		\hline
		$R$ [R$_{\oplus}$]             & 1.54$\pm$0.14                   & 2.29$\pm$0.26                    \\
		$i$ [deg]                      & 88.36$^{+0.17}_{-0.15}$         & 87.68$^{+0.31}_{-0.28}$          \\
		$a$ [AU]                       & 0.04994$^{+0.00048}_{-0.00049}$ & 0.09514$^{+0.00091}_{-0.00094}$  \\
		Insolation Flux [S$_{\oplus}$] & 426$^{+67}_{-60}$               & 117$^{+18}_{-16}$                \\
        $T_{\rm eq}$ [K] $^{(*)}$      & 1266$^{+44}_{-50}$              & 916$^{+32}_{-37}$                \\
		$M$ [M$_{\oplus}$]             & 7.3$^{+1.1}_{-1.0}$             & 8.3$^{+1.3}_{-1.3}$              \\
		$\rm \rho$ [g cm$^{-3}$]       & 11.0$^{+4.1}_{-2.8}$            & 3.8$^{+1.8}_{-1.1}$              \\
		\hline
	\end{tabular}
	\begin{minipage}{\columnwidth} 
	{\footnotesize $^{(*)}$ The equilibrium temperature values were calculated assuming null bond albedo.}
\end{minipage}	
\end{table}

Finally, we repeated our two-step analysis replacing the Keplerian models by sinusoidal functions to fit the planetary signals with null eccentricities and we obtained similar results. The RMS in the periodogram of the residuals after subtracting the two planets along with the long-term signal using the sinusoidal models (1.92~m~s$^{-1}$) is almost identical to the one obtained with the Keplerian approach (1.91~m~s$^{-1}$). Comparing $\log Z$ calculated from the posterior distribution of both models, we obtained values of $\log Z$~=~-103.79 for the sinusoidal fits, and $\log Z$~=~-103.38 for the Keplerian approach. The difference between the $\log Z$ values indicates that both models are equally favored according to Jeffreys's interpretation of the Bayesian factor \citep{Jeffreys61}, with slightly better results coming from the Keplerian model, which is supported by the marginally better RMS in the residuals.  

\section{Discussion} 

\label{sec:Discussion}

\subsection{Planet density and bulk composition}

\begin{figure}
	\includegraphics[width=\hsize]{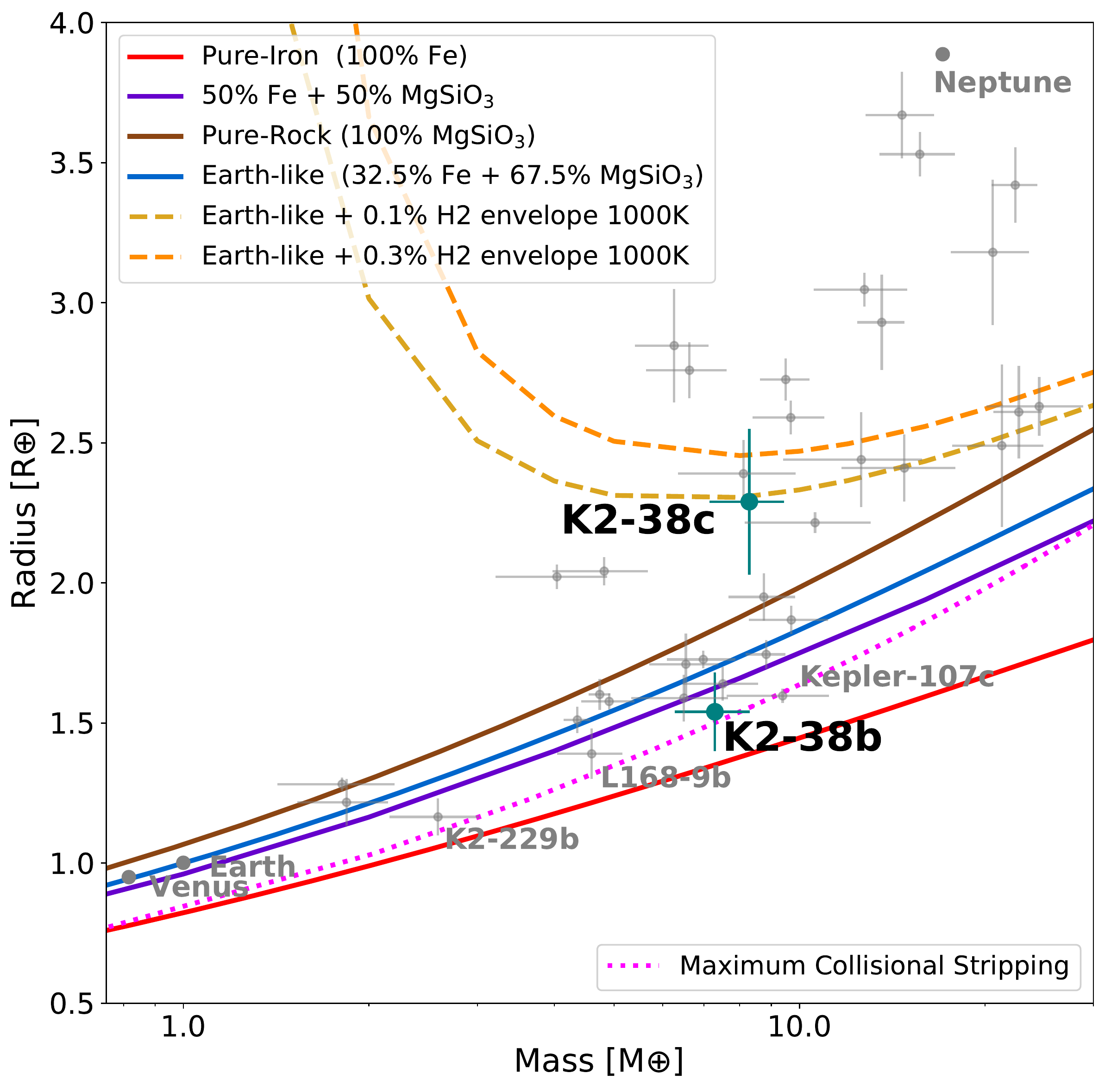}
    \caption{Radius-mass diagram including the planets from the NASA exoplanets archive with $>=$4 $\sigma$ parameter measurements and a public RV semi-amplitude measurement. Different models are plotted: the red one denotes a 100\%~Fe composition, the purple one denotes a 50\%~Fe~-~50\%~rock composition, the blue one denotes a 32.5\%~Fe~-~67.5\%~rock composition, the yellow and orange ones denote the same composition with a 0.1\% and 0.3\% mass percentage of H2 envelope, respectively, and the brown one denotes a rocky-type interior (composed by MgSiO$_{3}$). The dotted pink line indicates the minimum radius that a collision can produce for a certain planetary mass.}
    \label{fig:mass_radius}
\end{figure}

Our mass measurements point out to a different mass distribution with respect to those reported by \citet{2016ApJ...827...78S}, with K2-38c ($M_{\rm p}$=8.3$^{+1.3}_{-1.3}$~M$_{\oplus}$) being more massive than K2-38b ($M_{\rm p}$=7.3$^{+1.1}_{-1.0}$~M$_{\oplus}$). This difference comes from the time-span of the ESPRESSO dataset (three times larger), its smaller errobars (almost a factor two) and the larger number of measurements in comparison with the HIRES dataset. We calculated the density of both planets using the radius obtained in the photometric transit analysis and the masses coming from the spectroscopic MCMC analysis. We obtained mean densities of $\rho_{\rm p}$=11.0$^{+4.1}_{-2.8}$~g cm$^{-3}$ for K2-38b and $\rho_{\rm p}$=3.8$^{+1.8}_{-1.1}$~g cm$^{-3}$ for K2-38c. To compare these properties with the rest of the values published in the literature, we used the complete sample of confirmed planets from the NASA exoplanets archive. We selected planets with well-established (i.e., relative uncertainties below the 25\%) properties (orbital period, distance, inclination, radius, and mass) and with a published measure of the RV semi-amplitude (which ensures us to have mass measurements comparable with our results). We used this sample to create the mass-radius diagram shown in Fig.~\ref{fig:mass_radius} that includes four composition models from \citet{2013PASP..125..227Z} along with two additional models that include H2 envelopes from \citet{2019PNAS..116.9723Z}.

In Fig.~\ref{fig:mass_radius} we have represented different models that indicate the more probable compositions of the planets in the K2-38 system. K2-38b is better described by a composition of 50\%~Fe~-~50\%~rock but also consistent with models with higher percentages of Fe (we calculated a 67.6\% of Fe from the interior structure models computed through the wolfram tools developed by Li Zeng in \citealt{2013PASP..125..227Z} and \citealt{2016ApJ...819..127Z}). These models characterize iron-rich planets like Mercury in which the presence of a mantle and a planetary magnetic field is not common. This lack of mantle could be caused by collisions during the planetary formation \citep{2010ApJ...712L..73M}. To explore if this phenomenon is occurring in the K2-38 system, we represented in Fig.~\ref{fig:mass_radius} a collision-stripping boundary. This curve represents the minimum radius that a planet can have for a given mass, if it experienced typical collisional events throughout its history. The errorbars of K2-38b fall within the boundary, making the planet compatible with this scenario. We have highlighted in Fig.~\ref{fig:mass_radius} three objects with similar characteristics to K2-38b: Kepler-107c \citep{2019NatAs...3..416B}, L 168-9b \citep{2020A&A...636A..58A}, and K2-229b \citep{2018NatAs...2..393S}. The hydrodynamical simulations carried for Kepler-107c \citep{2019NatAs...3..416B} shown how the high density of the planet can be explained through the mantle stripping caused by giant collisions. The case of K2-229b is a good example of a Mercury analog Earth-sized planet with a very-short orbital period (14~hr) \citep{2018NatAs...2..393S}. The high metallicity of K2-38 could explain the formation of an iron-rich massive planet like K2-38b in this system. The theoretical models predict the formation of iron-dominated planets in close-in orbits \citep{2013ApJ...769...78W}. The most extreme case of this kind of planets is the planet KOI-1843.03, with an orbital period of only 4.25~hr, an estimated density $\rho_{\rm p}>$7~g cm$^{-3}$, and whose composition suggests a pure-iron structure \citep{2013ApJ...773L..15R}. In the case of K2-38c, the planet fits better within a model of 32.5\%~Fe~-~67.5\%~rock with a H2 atmosphere of 0.1\% by mass, although the errorbars make it compatible with the same envelope with a higher mass percentage (0.3\%).

\begin{figure}
	\includegraphics[width=\hsize]{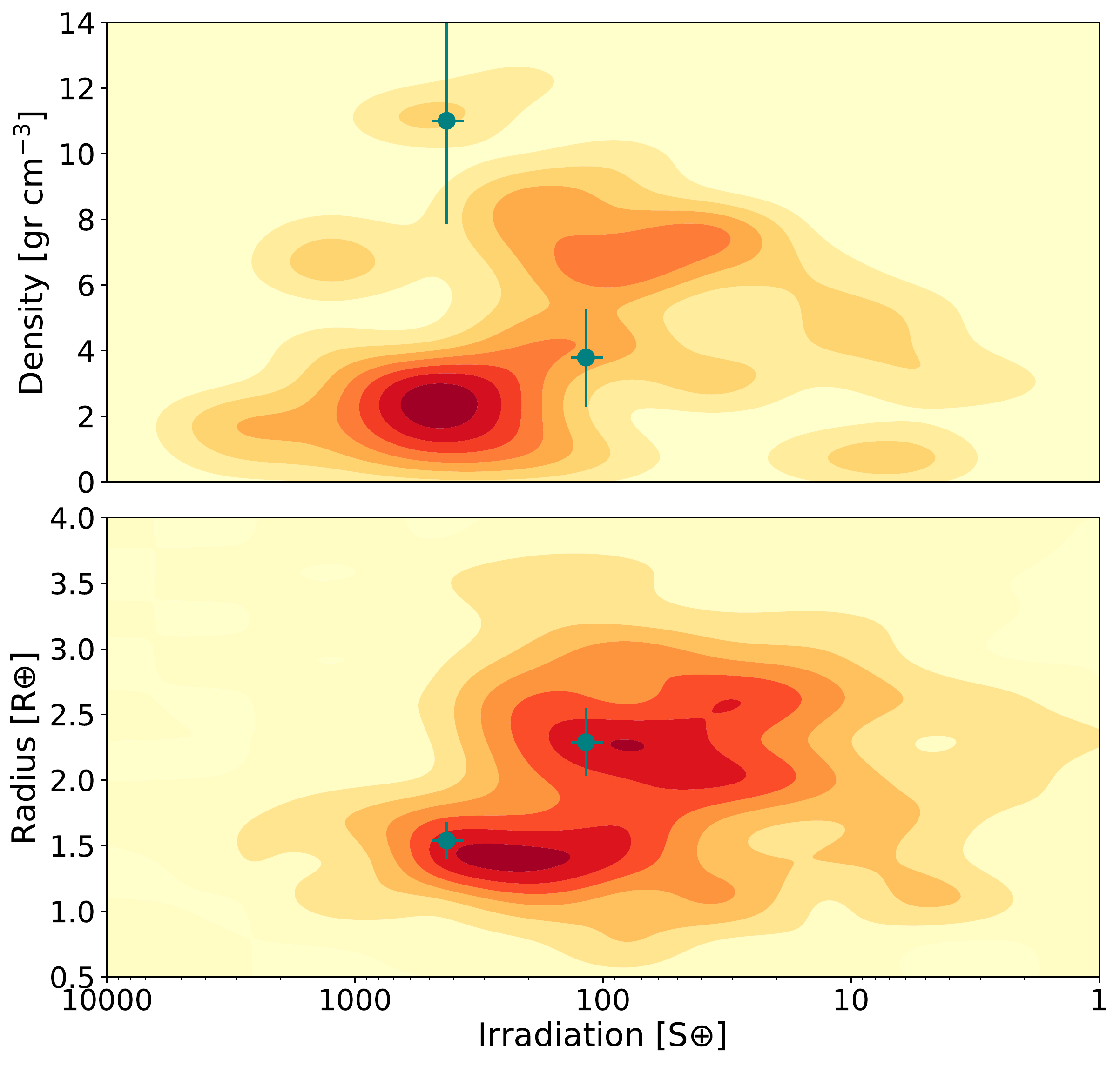}
    \caption{\textbf{Top:} Density-irradiance diagram with contours built using the exoplanets sample from Fig.~\ref{fig:mass_radius}. \textbf{Bottom:} Radius-irradiation diagram with contours built using an exoplanet sample from NASA archive that includes all exoplanets with known radius measurements.}
    \label{fig:irr_radius}
\end{figure}

To study the dependence of the density of these planets with the respective flux received from their host star, we calculated the incident flux from the parent star as a function of the effective temperature, the radius of the star, and the semi-major axis. We obtained a value of $S$=426$^{+67}_{-60}$~S$_{\oplus}$ for K2-38b and $S$=117$^{+18}_{-16}$~S$_{\oplus}$ for K2-38c. We calculated this quantity for two samples of exoplanets and we represented it with contours in Fig.~\ref{fig:irr_radius} against the planet density and the planet size. We created the contours by binning the data and interpolating using the \texttt{scipy.ndimage} package.

The top panel of Fig.~\ref{fig:irr_radius} includes contours built from the same exoplanets sample used in Fig.~\ref{fig:mass_radius}. This panel shows that K2-38b is among the densest planets detected to date, along with Kepler-107c ($\rho_{\rm p}$=12.7$\pm$2.5~g cm$^{-3}$, \citealt{2019NatAs...3..416B}), L 168-9b ($\rho_{\rm p}$=9.6$^{+2.4}_{-1.8}$~g cm$^{-3}$, \citealt{2020A&A...636A..58A}) and K2-229b ($\rho_{\rm p}$=8.9$\pm$2.1~g cm$^{-3}$, \citealt{2018NatAs...2..393S}). This can also be seen in Fig.~\ref{fig:mass_radius}, where we show that this planet is approximately twice as dense as the Earth. The mass distribution obtained by \citet{2016ApJ...827...78S} indicates that K2-38b could be the densest planet discovered, but our more precise results reveal that K2-38b is less massive than previously reported. In the case of K2-38c, its density is compatible with that of a sub-Neptune like planet. The difference in density of the two planets in spite of their similar masses may suggest that, despite sharing a similar original composition at the time when the planetary system was formed, the two planets have experienced an unrelated evolution due to differing stellar irradiation and possibly to their migration histories. As we can see by the maximum collision stripping represented in Fig.~\ref{fig:mass_radius}, the impacts suffered by K2-38b could be one of the main causes of this difference. The composition and the density of both planets are also related to the strength of their magnetic field \citep{2019MNRAS.490...15O}, with planets with iron-rich cores like K2-38b presenting higher densities and lower magnetic fields than the ice-rich ones.

The semi-major axis of both planets presented in Table~\ref{tab:K2-38_DerivedProperties} locates both planets in very close-in orbits (closer than Mercury in the Solar System), which indicates that they may be subject to evaporation effects (due to a higher insolation flux). The bottom panel of Fig.~\ref{fig:irr_radius} clearly shows each of these planets located on a different side of the radius valley. The largest one (K2-38c) is positioned above the gap, where the received flux is lower; while the smallest one (K2-38b) is located below, where the irradiance is higher. We estimated the mass-loss rate for both planets following the procedure described in \citet{2007A&A...461.1185L}. Using the semi-major axis, mass, and radius of the planets as input parameters, and assuming the UV luminosity of a G2 star, we obtained a mass loss of 0.058$\pm$0.024~M$_{\oplus}$ for K2-38b and 0.046$\pm$0.018~M$_{\oplus}$ for K2-38c along the life-time of the star. We included in the panel contours built from a sample of exoplanets whose parent stars have a published measurement of their radii. The radius gap of the valley \citep{2017AJ....154..109F} separates super-Earths ($\sim$1.5~R$_{\oplus}$) like K2-38b from sub-Neptunes ($\sim$2.5~R$_{\oplus}$) like K2-38c, and its slope has been already constrained in terms of the orbital period \citep{2018MNRAS.479.4786V}. This bimodal distribution has been found in 12 multi-planetary systems in a recent study carried out by \citet{2020MNRAS.491.5287O}, pointing out to the necessity of carrying out both spectroscopic and photometric follow-up to have a good characterization of the mass and the radius of the planets to define better the limits of this gap. In the case of a planetary system composed of a gaseous mini-Neptune and a rocky super-Earth (such as K2-38), this study also provides a formalism to calculate the minimum mass of the former in order for the system to be consistent within a photoevaporation scenario. In our case, we obtained a minimum mass of 2.49$^{+0.62}_{-0.64}$M$_{\oplus}$ for K2-38c, a threshold lower than the measured mass of this planet, which makes the K2-38 system consistent with the photoevaporation model. For the core-powered mass loss case, we relied on the formulation made by \citet{2020arXiv200301136C}. We computed a minimum mass of 7.73M$_{\oplus}$ for K2-38c, which is lower than our mass measurement, making this scenario also compatible with the masses obtained in our work.

The similarity in the mass of both planets suggests an evaporation threshold for small H/He envelopes between $\sim$150 and $\sim$400 times the irradiation of the Earth for a planetary-mass of $\sim$8~M$_{\oplus}$. This boundary will depend fundamentally on the spectral type of the star and also on its age. Different stellar environments will produce different evaporation rates \citep{2018AJ....156..264F}. In the irradiance-radius diagram of Fig.~\ref{fig:irr_radius} the lack of super-Earth at high irradiances found by \citet{2016NatCo...711201L} is also visible. They used a large sample of detected exoplanets from the Kepler mission, positioning the super-Earth gap between 2.2 and 3.8~R$_{\oplus}$ for irradiances higher than 650 times of the insolation flux of the Earth. In our case, only the region positioned below the radius valley related to smaller planets extends to irradiances higher than 650 times the irradiance of the Earth. 

\subsection{Co-orbital scenario} 

The high density found for the planet K2-38b makes it necessary to discard other possible scenarios that could equally explain the observed ESPRESSO radial velocity data. The co-orbital case is one of the main scenarios in which a two-planet system co-orbiting at the same stellar distance would wrongly increase the mass of a planet if the signal is interpreted as coming from a single Keplerian. Co-orbital configurations have not yet been confirmed outside of the Solar System although several candidates have recently been published (of particular interest is the case of TOI-178, \citealt{leleu19}; but see also \citealt{hippke15}, \citealt{janson13} or \citealt{lillo-box2018b}). Interestingly, co-orbital planet pairs are stable under a very relaxed condition developed in \cite{laughlin02} and such configuration would remain long-term stable as soon as the total mass of the planet and its co-orbital companion is smaller than 3.8\% of the mass of the star. This condition would be fulfilled by any co-orbital planetary-mass for K2-38.

Hence, we have explored this possibility by taking advantage of the transiting nature of the planet to apply the technique developed in \cite{leleu17} and subsequently applied in \cite{lillo-box2018a}. We modeled the radial velocity data by using Eq. 18 in \cite{leleu17}, where a new parameter $\alpha$ is included. This parameter is proportional to the mass ratio between the co-orbital and the main planet. Hence, if compatible with zero, we can discard co-orbitals to a certain planetary mass. We performed this test using the \texttt{emcee} package with 50 walkers and 10\,000 steps per walker to explore the parameter space, and in particular to explore the posterior distribution of the $\alpha$ parameter. The model includes the aforementioned equation for planet K2-38b and a simple Keplerian signal for K2-38c. The result of this test provides an $\alpha$ posterior characterized by $\alpha=0.42^{+0.37}_{-0.44}$. The shift of the posterior distribution with respect to zero is not statistically significant given its uncertainty. However, the large separation from the null value ($\sim 1\sigma$) makes impossible to reach a firm conclusion on the possible presence of co-orbitals. Additional data is thus required to narrow down the posterior distribution in order to provide further constraints on the co-orbital scenario. From the RMS of the out-of-transit light curve, we estimated an upper limit on the co-orbital radius of $R_{\rm co}$=1.01$\pm$0.07~R$_{\oplus}$ assuming a coplanar orbit. As a side note, if we would have measured a value $\alpha=0.42$, this would mean a reduction in the mass of the main planet due to the mass repartition between the two co-orbitals of around 30\%, placing the planet K2-38\,b in the Earth-like density regime.

\subsection{Additional RV signals and stellar rotation} 

\begin{figure}
	\includegraphics[width=\hsize]{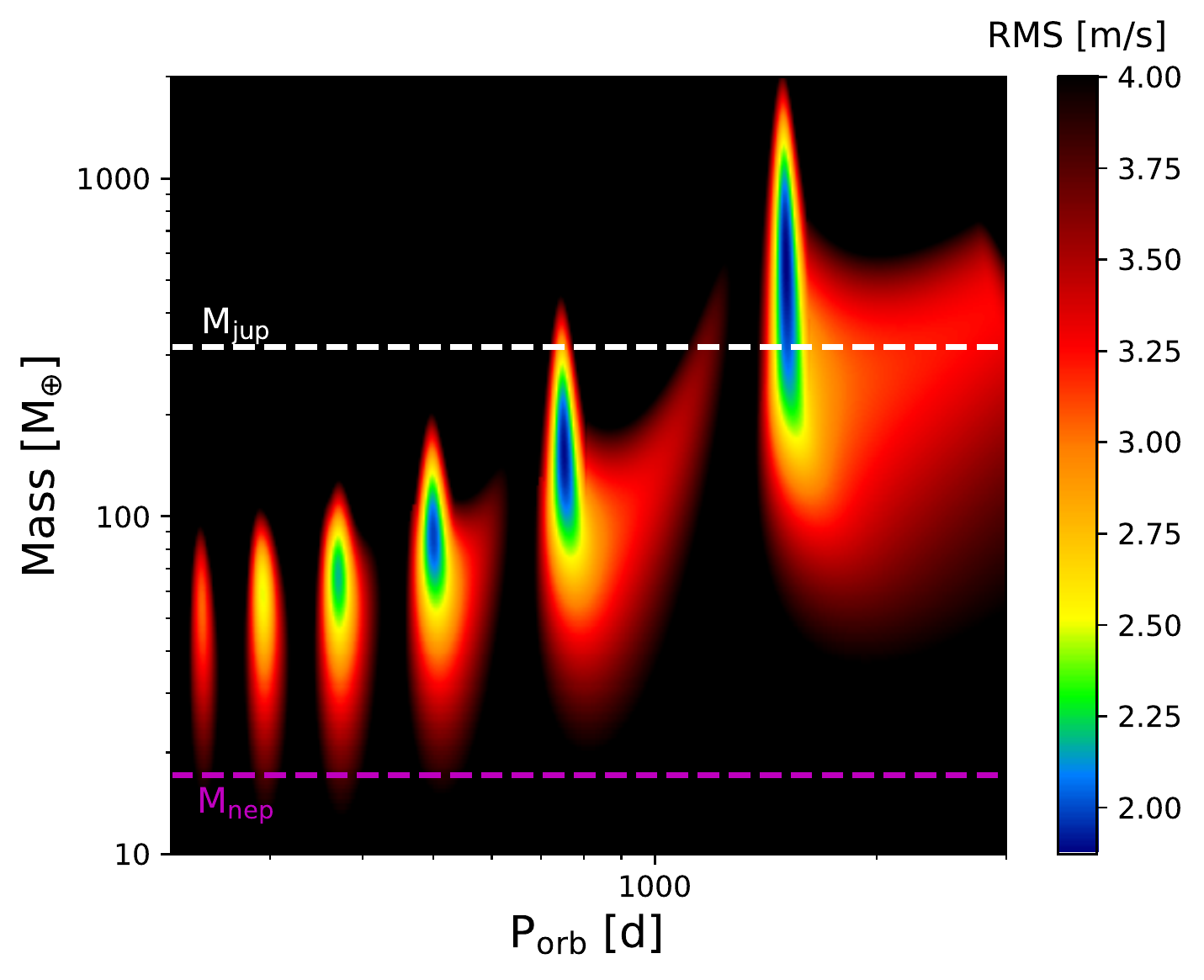}
    \caption{Distribution of the RMS of the residuals in the RV-time series after subtracting the long-term signal with a sinusoidal whose amplitude is calculated from the considered values of mass and period. The horizontal white and purple dashed lines represent the Jupiter and Neptune mass, respectively.}
    \label{fig:chi2}
\end{figure}

Despite the fact of not finding more significant signals in the MCMC RV analysis, we added a third Keplerian fit to our model to search a possible third planet. We included the new five parameters without bounds using 30\,000 steps for the burn-in stage and 150,000 for the construction stage, but the MCMC simulations did not converge to a clear result. We then considered possible planetary solutions that can be causing the long-term signal in the RV measurements, a possibility that \citet{2016ApJ...827...78S} explored through an analysis of the RV acceleration using the Bayesian Information Criterion (BIC) showing that the possibility of a third companion is feasible but without providing a clear mass limit. A potential companion around K2-38 have been reported by \citet{2018A&A...610A..20E} and \citet{2020A&A...635A..73B}. In the first case, the companion was identified by the Gaia DR2 as a background object. In the second case, the probability of this companion to be a background object is only 1.59\%, but more astrometric measurements are required to check the common proper motion. We used the offset-corrected RV time-series after the subtraction of the two-planet model, and model the long-period signal using a sinusoidal with only the phase as a free parameter along with a constant. We calculated the amplitude of this sinusoidal using Eq.~\ref{eq:Mass} for a certain range of masses and periods. We then computed the RMS of the residuals for all the models and represent it in the colormap of Fig.~\ref{fig:chi2}.

From Fig~\ref{fig:chi2} one can conclude that the period of this long-term signal must be within the regions of ~375, 750 or 1500d (where the residuals are minimized under a 2~m/s threshold). If this signal is caused by a planet, this would have a mass between 0.25 and 3~M$_{\rm J}$. With a longer ESPRESSO RV time-series, we would be able to constrain better this signal, along with a more accurate model for the two already detected planets which will produce changes in the distribution shown in Fig.~\ref{fig:chi2}. The Pearson coefficient \citep{1895RSPS...58..240P} calculated between the RV of the long-period signal and the stellar activity time-series of FWHM and NaD show a possible correlation between them. This indicates that the chromospheric activity of the star could be causing this signal, although further observations are necessary to confirm it and discard a possible planetary origin.

The fact that we could not find any hint of rotation neither in photometry and spectroscopy could indicate that the star may be in a Magnetic Grand Minimum (MGM) of stellar activity \citep{2012IAUS..286..335S}. If this would be the case, we expect a low amplitude for the amplitude of the cycle signal, which favors the interpretation of the long-period signal as a planet. Using the relation between the level of chromospheric activity and the RV semi-amplitude induced by the rotation of the star found by \citet{2017MNRAS.468.4772S} we estimated an induced RV semi-amplitude lower than 0.60~m~s$^{-1}$. This value is lower than the RMS of the residuals after subtracting the two planets and the long-term signal, which indicates that it is not possible to detect the rotation signal in the RV time-series until we have more measurements. This lack of detectability comes mainly from the low chromospheric activity of the star along with its age (older than the Sun) and intrinsic RV errors (around 1~m~s$^{-1}$).

\section{Conclusions}

\label{sec:Conclusions}

From the analysis of the K2 photometric light curve, we have measured the radius of the two planets transiting the star K2-38. We characterized K2-38b as a super-Earth with $R_{\rm p}$=1.54$\pm$0.14~R$_{\oplus}$, and K2-38c as a sub-Neptune with $R_{\rm p}$=2.29$\pm$0.26~R$_{\oplus}$. Both radii are in good agreement with the values previously reported in the literature. From the RV analysis, we have used a two-planet model based on Keplerians with offset and trend correction (this last one made with a sinusoidal fit) to obtain the rest of the planetary parameters. The MCMC simulations produced a RMS of 1.91~m~s$^{-1}$ for the combined HIRES and ESPRESSO dataset, providing a mass of $M_{\rm p}$=7.3$^{+1.1}_{-1.0}$~M$_{\oplus}$ for K2-38b and $M_{\rm p}$=8.3$^{+1.3}_{-1.3}$~M$_{\oplus}$ for K2-38c. These results shows a significant reduction in the uncertainties with the respect to the previous mass measurements published in the literature thanks to the higher RV precision of ESPRESSO. In the case of K2-38b we measured a lower mass, difering 2$\sigma$ with respect to previous results, while in the case of K2-38c our result is compatible with the previous mass measurement. We did not find any significant signal in the RV residuals after subtracting the final model, although the long-term signal shows compatibility with a third planet of $M_{\rm p}$=0.25-3~M$_{\rm J}$, but could also have a chromospheric activity origin.

The expected rotation signal of the star ($P_{\rm rot}$=29~$\pm$~2d) was not detected either in the photometric light curves or in the RV time-series due to its low amplitude. None of the four spectroscopic indexes show any hint of this signal. We predict an upper limit to the induced RV semi-amplitude associated with rotation of 0.60~m~s$^{-1}$, making it challenging to detect this signal in the RV time-series.

We derived the mean density of each planet at $\rho_{\rm p}=$11.0$^{+4.1}_{-2.8}$~g cm$^{-3}$ for K2-38b and $\rho_{\rm p}=$3.8$^{+1.8}_{-1.1}$~g cm$^{-3}$ for K2-38c, confirming K2-38b as one of the densest planets known to date. The best model for the composition of this planet comes from an iron-rich Mercury-like model, while K2-38c is better described by a rocky model with a H2 envelope. The high density of K2-38b could be explained through the mantle stripping theory based on giants collisions. According to their derived irradiances, each planet is located on a different side of the radius valley, resulting from the different irradiation levels and evaporation processes at which they are exposed, along with core-powered mass loss mechanisms.

\begin{acknowledgements}

B.T.P., A.S.M., J.I.G.H., R.R., C.A.P. acknowledge financial support from the Spanish Ministry of Science and Innovation (MICINN) project AYA2017-86389-P. B.T.P. acknowledges Fundaci\'on La Caixa for the financial support received in the form of a Ph.D. contract. C.L. and F.A.P. would like to acknowledge the Swiss National Science Foundation (SNSF) for supporting research with ESPRESSO through the SNSF grants nr. 140649, 152721, 166227 and 184618. The ESPRESSO Instrument Project was partially funded through SNSF’s FLARE Programme for large infrastructures. S.C.C.B. acknowledges support from FCT through Investigador FCT contract IF/01312/2014/CP1215/CT0004. J.I.G.H. acknowledges financial support from the Spanish MICINN under the 2013 Ram\'on y Cajal program RYC-2013-14875. This work was supported by FCT - Fundação para a Ciência e a Tecnologia through national funds and by FEDER through COMPETE2020 - Programa Operacional Competitividade e Internacionalização by these grants: UID/FIS/04434/2019; UIDB/04434/2020; UIDP/04434/2020; PTDC/FIS-AST/32113/2017 \& POCI-01-0145-FEDER-032113; PTDC/FIS-AST/28953/2017 \& POCI-01-0145-FEDER-028953; PTDC/FIS-AST/28987/2017 \& POCI-01-0145-FEDER-028987; PTDC/FIS-OUT/29048/2017. V.Z.A., S.G.S., S.C.C.B., N.J.N. acknowledge support from FCT through Investigador FCT contracts nsº IF/00650/2015/CP1273/CT0001; IF/00028/2014/CP1215/CT0002; IF/01312/2014/CP1215/CT0004; IF/00852/2015. S.G.S., N.J.N. also acknowledge support from FCT in the form of an exploratory project with the reference IF/00028/2014/CP1215/CT0002; IF/00852/2015. J.P.F., O.D. acknowledge support from FCT through national funds in the form of a work contract with the references DL 57/2016/CP1364/CT0005; DL 57/2016/CP1364/CT0004. This work has been carried out within the framework of the National Centre of Competence in Research PlanetS supported by the SNSF. C.L., F.B., R.A., B.L., F.A.P., D.E. \& S.U. acknowledge the financial support of the SNSF. This project has received funding from the European Research Council (ERC) under the European Union’s Horizon 2020 research and innovation programme (project {\sc Four Aces}; grant agreement No 724427). This paper includes data collected by the Kepler mission. Funding for the Kepler mission is provided by the NASA Science Mission Directorate. This work makes use of The Data Analysis Center for Exoplanets (DACE), which is a facility based at the University of Geneva (CH) dedicated to extrasolar planets data visualization, exchange, and analysis. DACE is a platform of the Swiss National Centre of Competence in Research (NCCR) PlanetS, federating the Swiss expertise in Exoplanet research. The DACE platform is available at https://dace.unige.ch. This research has made use of the NASA Exoplanet Archive, which is operated by the California Institute of Technology, under contract with the National Aeronautics and Space Administration under the Exoplanet Exploration Program.

\end{acknowledgements}

%
   \bibliographystyle{aa} 
   \bibliography{References} 
%

\section*{Affiliations}

\noindent $^{\boldsymbol{1}}$ Instituto de Astrof\'isica de Canarias, E-38205 La Laguna, Tenerife, Spain

\noindent $^{\boldsymbol{2}}$ Universidad de La Laguna, Departamento de Astrof\'isica, E-38206 La Laguna, Tenerife, Spain

\noindent $^{\boldsymbol{3}}$ Observatoire Astronomique de l'Universit\'e de Gen\`eve, 51 chemin des Maillettes, 1290 Versoix, Switzerland

\noindent $^{\boldsymbol{4}}$ Instituto de Astrofísica e Ciências do Espaço, Universidade do Porto, CAUP, Rua das Estrelas, 4150-762 Porto, Portugal 

\noindent $^{\boldsymbol{5}}$ Departamento de Física e Astronomia, Faculdade de Ciências, Universidade do Porto, Rua do Campo Alegre, 4169-007 Porto, Portugal 

\noindent $^{\boldsymbol{6}}$ INAF Osservatorio Astrofisico di Torino, Via Osservatorio 20, 10025 Pino Torinese, Italy 

\noindent $^{\boldsymbol{7}}$ Centro de Astrobiología (CSIC-INTA), Carretera de Ajalvir km 4, 28850 Torrejón de Ardoz, Madrid, Spain

\noindent $^{\boldsymbol{8}}$ Consejo Superior de Investigaciones Cient\'ificas, E-28006 Madrid, Spain

\noindent $^{\boldsymbol{9}}$ INAF Osservatorio Astronomico di Trieste, via G. Tiepolo 11, 34143 Trieste, Italy

\noindent $^{\boldsymbol{10}}$ Centro de Astrobiología (CAB, CSIC-INTA), Dpto. de Astrofísica, ESAC campus 28692 Villanueva de la Cañada (Madrid), Spain

\noindent $^{\boldsymbol{11}}$ Scuola Normale Superiore, Piazza dei Cavalieri, 7, 56126 Pisa, Italy

\noindent $^{\boldsymbol{12}}$ European Southern Observatory, Alonso de Córdova 3107, Vitacura Casilla 19001, Santiago 19, Chile

\noindent $^{\boldsymbol{13}}$ European Southern Observatory, Karl-Schwarzschild-Str. 2, 85748 Garching bei München, Germany 

\noindent $^{\boldsymbol{14}}$ INAF Osservatorio Astronomico di Palermo, Piazza del Parlamento 1, 90134 Palermo, Italy

\noindent $^{\boldsymbol{15}}$ Instituto de Astrofísica e Ciências do Espaço, Universidade de Lisboa, Edifício C8, 1749-016 Lisboa, Portugal

\noindent $^{\boldsymbol{16}}$ Departamento de Física da Faculdade de Ciências da Universidade de Lisboa, Edifício C8, 1749-016 Lisboa, Portugal

\noindent $^{\boldsymbol{17}}$ Physikalisches Institut \& Center for Space and Habitability, Universität Bern, Gesellschaftsstrasse 6, 3012 Bern, Switzerland

\noindent $^{\boldsymbol{18}}$ INAF Osservatorio Astronomico di Brera, Via E. Bianchi 46, 23807 Merate, Italy

\noindent $^{\boldsymbol{19}}$ INAF Osservatorio Astronomico di Padova, Vicolo dell’Osservatorio 5, I-35122, Padova, Italy

\noindent $^{\boldsymbol{20}}$ Centro de Astrofísica da Universidade do Porto, Rua das Estrelas, 4150-762 Porto, Portugal

\noindent $^{\boldsymbol{21}}$ Institute for Fundamental Physics of the Universe, Via Beirut 2, 34151 Miramare, Trieste, Italy

\noindent $^{\boldsymbol{22}}$ Fundación Galileo Galilei, INAF, Rambla José Ana Fern2ndez Pérez 7, 38712 Breña Baja, Spain

\begin{appendix} 

\section{MCMC analysis}

\begin{figure*}
	\includegraphics[width=17.5cm,page=4]{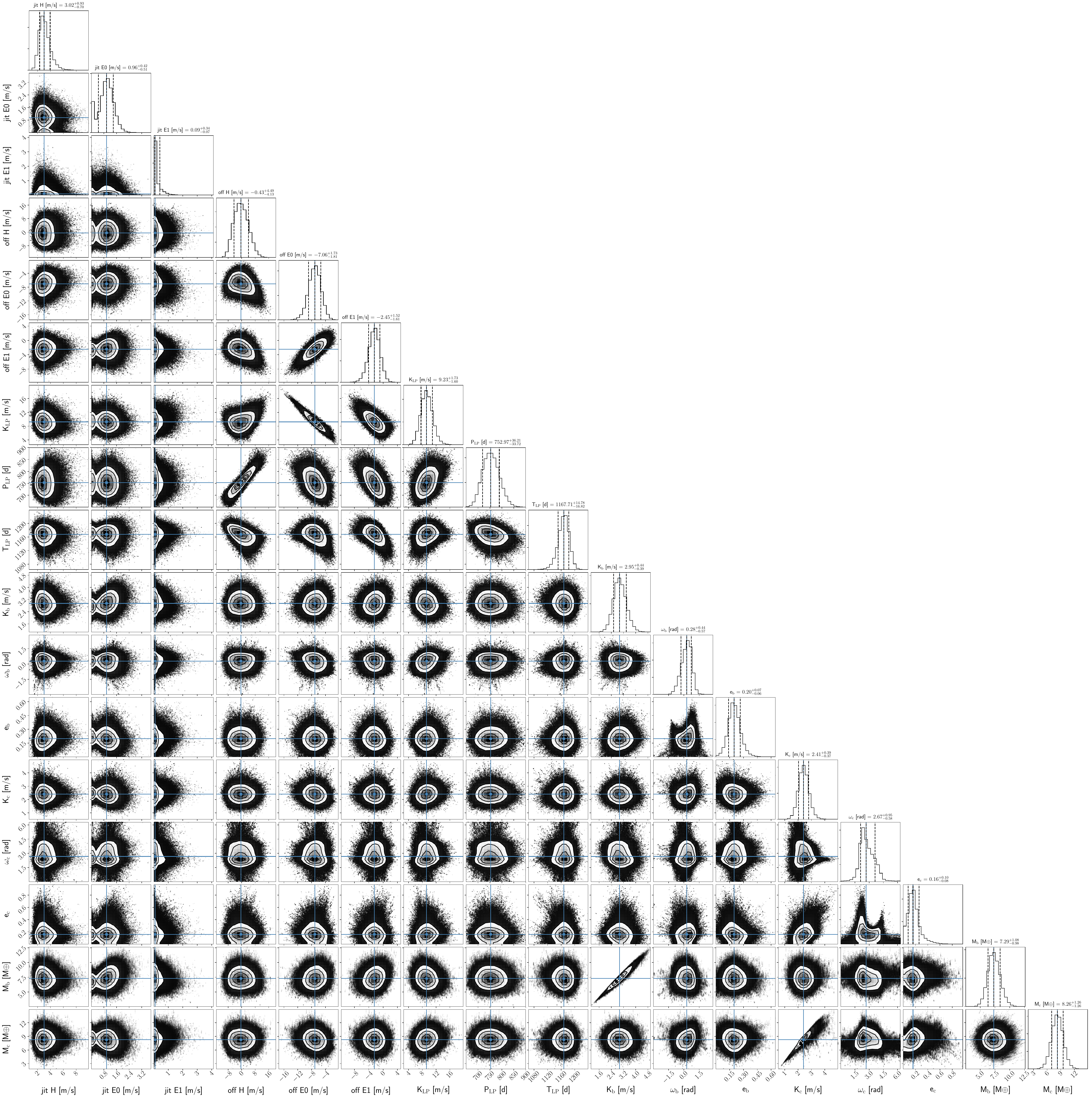} 
    \caption{Corner plot of all the fitted parameters included in the 2 planets model for the K2-38 RV time-series. The vertical lines indicate the mean and the 16th-84th percentiles.}
    \label{fig:RV_MCMC}
\end{figure*}

\begin{figure*}
	\includegraphics[width=17.5cm,page=1]{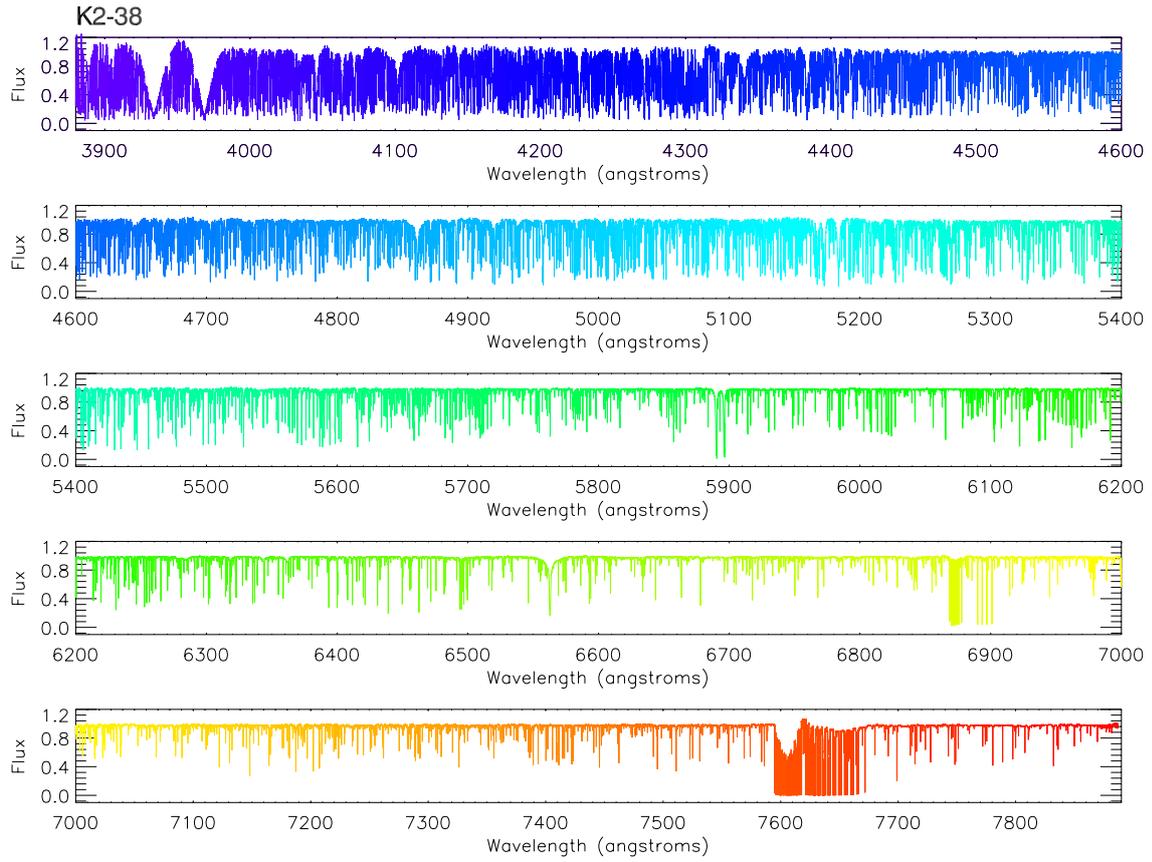} 
    \caption{Normalized 1D ESPRESSO spectrum of K2-38 corrected from RV. Five different spectral regions are shown in different colors.}
    \label{fig:K2-38_Spectrum}
\end{figure*}

\end{appendix}

\end{document}